**SMART OPERATORS IN INDUSTRY 4.0: A HUMAN-CENTERED APPROACH TO ENHANCE OPERATORS' CAPABILITIES AND COMPETENCIES WITHIN THE NEW SMART FACTORY CONTEXT**

Francesco Longo[a], Letizia Nicoletti[b], Antonio Padovano[c]
[a]DIMEG, University of Calabria, Via P. Bucci, Cube 45C, third floor, 87036 Rende, CS, Italy, f.longo@unical.it
[b]CAL-TEK Srl, Via Spagna 240-242, 87036 Rende, Italy
[c]DIMEG, University of Calabria, Via P. Bucci, Cube 45C, third floor, 87036 Rende, CS, Italy, antonio.padovano@unical.it

## ABSTRACT

As the Industry 4.0 takes shape, human operators experience an increased complexity of their daily tasks: they are required to be highly flexible and to demonstrate adaptive capabilities in a very dynamic working environment. It calls for tools that could be easily embedded into everyday practices and able to combine complex methodologies with high usability requirements. In this perspective, the proposed research work is focused on the design and development of a practical solution, called Sophos-MS, able to integrate augmented reality contents and intelligent tutoring systems with cutting-edge fruition technologies for operators' support in complex man-machine interactions. After establishing a reference methodological framework for the smart operator concept as part of the Industry 4.0 paradigm, the proposed solution is presented and functional and non-function requirements are discussed. Such requirements are fulfilled through a structured design strategy whose main outcomes include a multi-layered modular solution, Sophos-MS, that relies on Augmented Reality contents and on an intelligent personal digital assistant with vocal interaction capabilities. The proposed solution has been deployed and its potentials as a training tool have been investigated with field experiments. The experimental campaign results have been firstly checked to ensure their statistical relevance and then analytically assessed in order to show that the proposed solution has a real impact on operators' learning curves and can make the difference between who uses it and who does not.

**Keywords: smart factory, industry 4.0, augmented reality, smart operators, intelligent vocal assistance**

## INTRODUCTION

When dealing with the most recent developments in manufacturing systems, smart factory and industry 4.0 are crucial keywords. Such concepts are meant to capture the latest trends as well as the key requirements for gaining sustainable competitive advantages in the global arena. The capabilities of digital solutions/tools have opened up new opportunities and raised ambitious challenges for manufacturing systems. As a matter of facts, while interoperability, information transparency, technical assistance and decentralized decisions stand out as essential design principles for operational and management efficiency, it is crucial to detect and achieve an optimum integration level between physical and virtual reality. It goes with intelligent networks of Cyber-Physical Systems and Human Resources communicating over the Internet of Things and the Internet of Services. Thus, smart factories are featured by a seamless integration of advanced manufacturing capabilities with digital infrastructures able to capture, generate and spread intelligence through improved monitoring, analytics, modeling and simulation. All these aspects throw traditional production and management paradigms aside and call for a complete overhaul of businesses, procedures and structures. Hence, notwithstanding the Industry 4.0 concept is well conceptualized and clear in its foundations, it is difficult to achieve due to systems diversity and given the multiplicity of available/conceivable solutions. As a matter of facts, a state of the art review of Industry 4.0 based on research and practice can be found in Weyer et al. (2015) and Stock and Selinger (2016) but, as highlighted in Qin et al. (2016), the technology roadmap for accomplishing Industry 4.0 is still not clear and the gap analysis between current manufacturing systems and Industry 4.0 requirements shows that there is still a long way to go.

To this end, Quin et al. (2016) propose a categorical, hierarchical framework where Industry 4.0 is achievable through a continuous and incremental process of evolution whose main dimensions are automation and intelligence: the intelligent manufacturing system is highly automated, at factory level, and self-aware, self-optimizing and self configuring. This framework can be easily and profitably coupled with the 5C architecture for Cyber Physical Systems in Industry 4.0 proposed in Lee et al. (2015) and Bagheri et al. (2015) as sketched out in Figure 1.

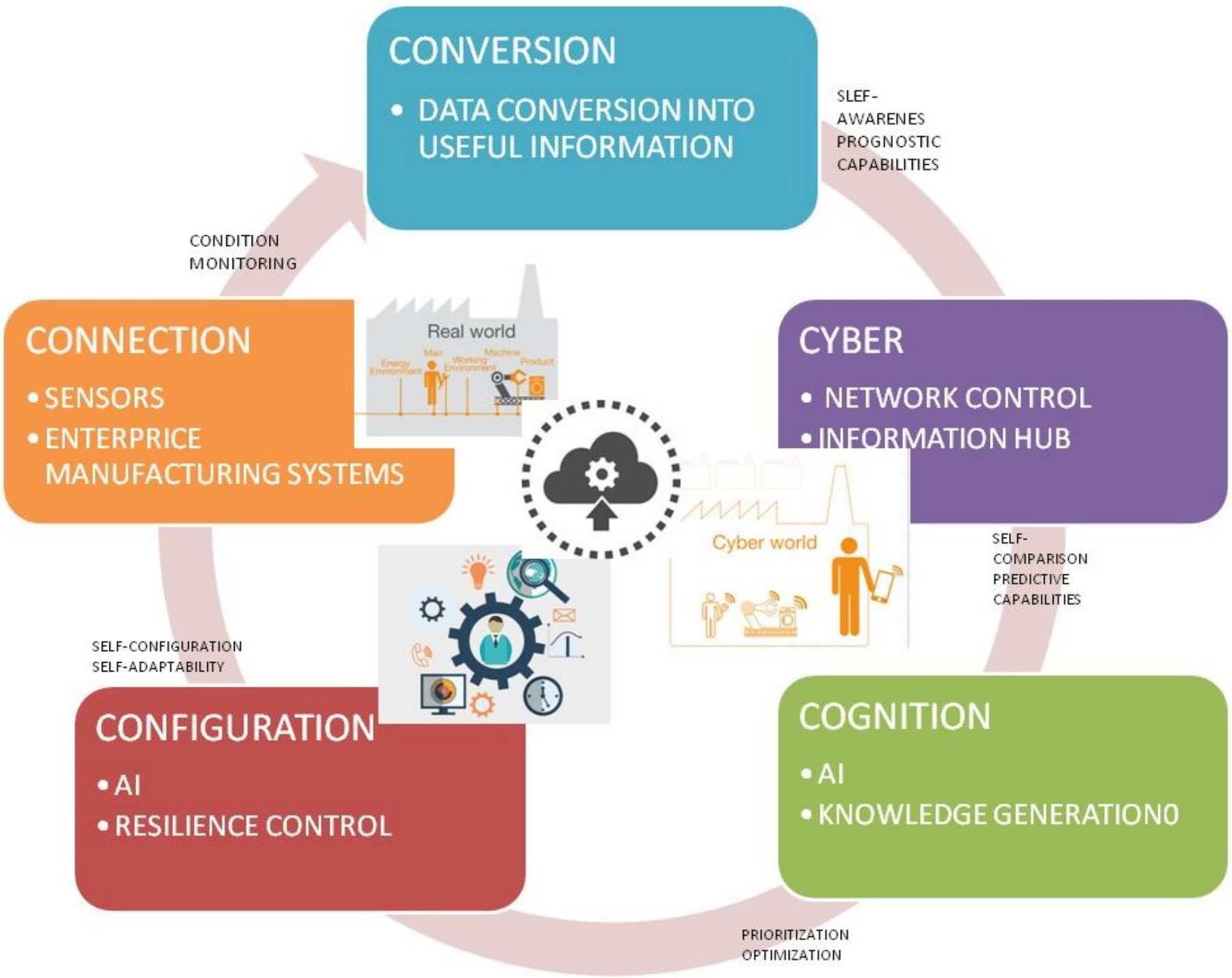

**Figure 1: 5C Architecture for Industry 4.0**

Thus, it is envisaged that the migration toward the factory of the future would require a structured approach lead by an organic and forward-looking vision where the Smart Product, the Smart Machine and the Augmented Operator are central paradigms (Weyer et al., 2015). Within the scope of the present study, particular focus is paid on the Augmented Operator paradigm. As the Industry 4.0 takes shape, indeed, human operators experience an increased complexity of their daily tasks: they are required to be highly flexible and to demonstrate adaptive capabilities in a very dynamic working environment. Therefore this research work is aimed at introducing a visionary and user-centered solution that could profitably work in the general framework of Industry 4.0.

## 2. Literary background

Research contributions to Industry 4.0 mainly include theoretical insights and conceptualization efforts to explore the many facets related to the Industry 4.0 paradigm. Success factors for the profitable introduction of hyper-connected smart factory are investigated in Park (2016). Human factors, in particular, are extensively considered in Kaare and Otto (2015) where a human centric employee performance measurement system is proposed based on a set of parameters gathered from medical and publicly available sensors and smart phones.

Moreover, as the inner complexity of manufacturing systems grows, proper workforce qualification strategies are required. To this end, an holistic competence model can be found in Hecklau et al. (2016). The proposed model is based on competencies identification, aggregation, categorization and visualization through radar charts.

As for smart manufacturing applications, a multi-agent approach can be found in Fedorov et al. (2015). It includes some basic building blocks of an automation system such as channel selection, decision making and storage. To enable research and development, instead, Helu and Hedberg (2015), design and architect a product lifecycle test bed featured by the integration of heterogeneous solutions including tools for enterprise resource planning (ERP), manufacturing execution systems (MES), and quality management systems (QMS). However, the technologies within the test bed infrastructure are not implemented and the definition of standards for interoperability among systems and applications remains an open issue. In this context, the SmartFactory$^{KL}$ has realized a very first multi-vendor and highly modular production system as a sample reference for Industry 4.0 (Weyer et al. 2015). In such a heterogeneous environment the necessary standards (mechanical, electrical and communication standards) have been defined to guarantee a smooth interaction among all the involved systems. On the other hand, another effort to make more tangible the vision of Industry 4.0 can be attributed to Erol et al. (2016). Although still in a planning stage, Erol et al. (2016) devise the TU Wien Industry 4.0 Pilot Factory where a learning Factory approach and scenario-based learning are combined to: promote research in the smart production, involve SMEs without own research infrastructures as test-beds, support SMEs needing special technical competencies, create application-oriented competencies among students. Another interesting contribution can be found in Jung et al. (2016) where, based on both research and practical experiences in the manufacturing area, a reference activity model for smart factory design and improvement is presented. For manufacturing enterprises maturity assessment, a quite general and comprehensive approach can be found in Schumacher et al (2016). Here, the maturity assessment model covers a variety of dimensions (Governance, Technology, Leadership, Products, Customers, Operations, Culture, People) and is coded in a practical computation tool that greatly enhances its application potentials both as a self assessment tool for further strategic measures and as a mean to collect data on the state of development across different industries to identify success factors for effective Industry 4.0 strategies.

Anyway, it goes without saying that the viability and success of any implementation initiative toward smart manufacturing requires a careful analysis of enabling technologies and implementation barriers. It calls for a proper risk assessment framework that allows manufacturers and solution providers to identify areas of higher risk as well as the primary weaknesses of a technology within the context of a particular manufacturing system. To this end Helu et al. (2015) introduce an Implementation Risk Assessment Framework (IRAF). Relevant side effects of Industry 4.0 have also been explored. For instance, Ivanov et al. (2015) develop a dynamic scheduling environment for short-term scheduling in smart factories. Moreover an IoT-based production performance model can be found in Hwang et al. (2016). The model is validated in a virtual factory and highlights the possibility of assessing Overall Equipment Effectiveness in real time.

The state of the art analysis shows that industry 4.0 is still an open research field were much has been done but there is still more to do to accomplish its vision. With this in mind, the proposed research builds upon previous contributions with particular reference to Lee et al. (2015), Bagheri et al. (2015) and Helu et al. (2015) and proposes a forward-looking implementation framework for the augmented operator paradigm as part of the industry 4.0 vision. Indeed, as previously discussed, there has been a great deal of efforts toward smart factory concepts and engineering. However, such efforts are mostly related to automation systems, plant solutions, communication infrastructures, systems connectivity and interoperability, data flows management, etc. Taking advantage of previous achievements, this research is intended to take a step forward proposing a human-centered approach along with its implementation and deployment to align (and enhance) operators' capabilities/competencies with the new smart factory context. The main research contribution is twofold: (i) a methodological framework aimed at shaping the augmented operator paradigm within the industry 4.0 vision; (ii) a breakthrough solution that implements the aforementioned framework .

### 3. Smart Operators in Smart Factories: a methodological framework

The proposed framework combines well-established methods and new approaches for operators' support in manufacturing systems. Starting from the 5C architecture depicted in figure 1 and presented in Lee et al. (2015) and Bagheri et al. (2015), the proposed framework goes over each building block considering how to enhance and/or integrate each in order to align operators' capacities/means with new requirements arising from smart factory context. In other words, the proposed framework offers an integrative approach, encompassing health, technical and organizational aspects, aimed at enabling human resources to be profitably part of the industry 4.0 overall design. To this end, human factors become the gravity center of the smart manufacturing Cyber Physical Space (CPS). Smart manufacturing entails an intelligent network where the physical context is closely intertwined with a corresponding cyber twin by means of IoT and Cloud Computing Infrastructures. Moreover, the cyber twin context is meant to be provided with intelligence to enable self-awareness, self-comparison, self-management and self-adaptivity at operational and procedural level. Hence the physical and the cyber contexts are meant to be seamlessly integrated so as to act synergistically. In this perspective, operators are a crucial link for the optimal integration between real and virtual. Human factors, indeed, are meant to benefit from the intelligence generated within the cyber context and in turn to add further intelligence, a meta-intelligence level, generating a virtuous closed-loop chain with a valuable feedback system that makes the overall manufacturing system grow and evolve over time toward even greater levels of efficacy and efficiency with an optimal integration of quantitative metrics and qualitative factors ensured by the synergic cooperation of human and artificial intelligence in a context aware CPS. In this perspective, a crucial aspect is in the interaction patterns (mean and resources) that involve human resources and physical contexts as well as human resources and the twin cyber context. To sort it out the augmented operator paradigm needs to be defined. The operator is augmented because of the capability to interact with intangible assets and digital contents in highly interactive as well as absorbing fruition experiences. In other words, his ability to perceive and act within the physical world is enhanced by the possibility to be immersed in a virtual reality environment where different contents levels superimpose each other. Moreover, Augmented Reality (AR) applications are suited to connect virtual and real objects since AR takes the current view of the real world and adds digital resources on top of it. So the augmented operator owns a superior knowledge of the working environment deriving not only from daily interactions due to operational tasks/ procedures but even with a variety of value added contents that are suited to augment his skills and abilities to perceive and act within the working environment. Such contents include: augmented reality, virtual reality, 3D reconstructions as well as digital contents of any type. Levels of immersion are related to the technology used to deliver such contents. A representation of this framework is provided in Figure 2.

## SOPHOS-MS Framework

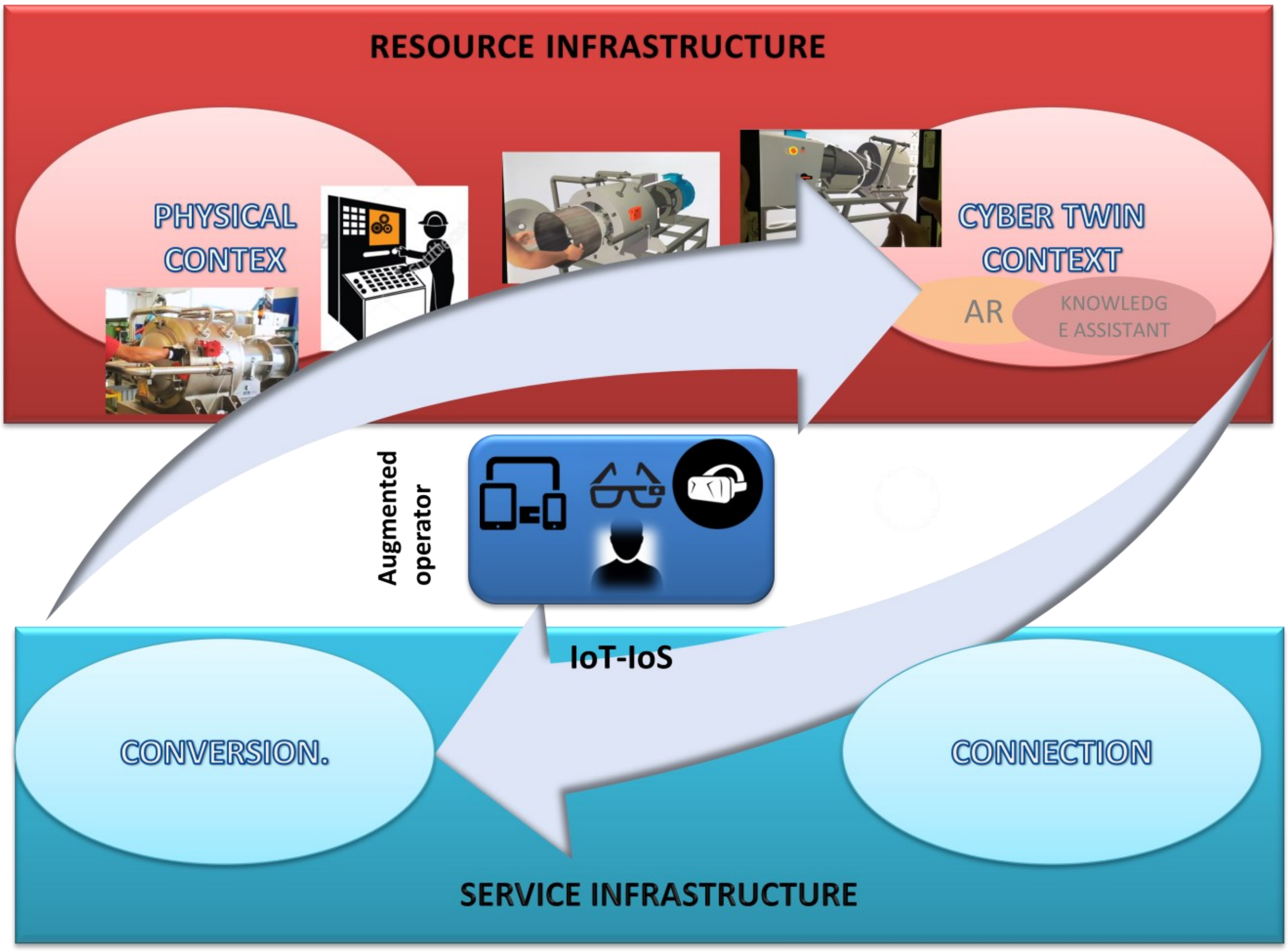

Figure 2: SOPHOS-MS Framework

For instance great levels of immersion can be achieved through headsets that allow users to look around, or even walk, in a virtual environment and, as they look/move, their view of the virtual world adjusts the same way it would if they were looking or moving in "real" reality. On the other hand, when it is required to leave user's hands free for tasks execution, smart glasses could be a good option.

In addition, aside from already existing approaches and technologies (like those mentioned above), the proposed framework conceptualizes and designs a new approach/component/solution to extend operators' capabilities within the smart factory paradigm. It is an intelligent personal digital assistant with vocal interaction capabilities and therefore able to answer operator's questions about tasks/procedures/equipments. It is meant to provide operators with quick and effective support allowing them to acquire the information they need through a Q&A approach as they would if they were talking with a knowledgeable expert. The SOPHOS-MS architecture concerning the personal digital assistant is given in Figure 3.

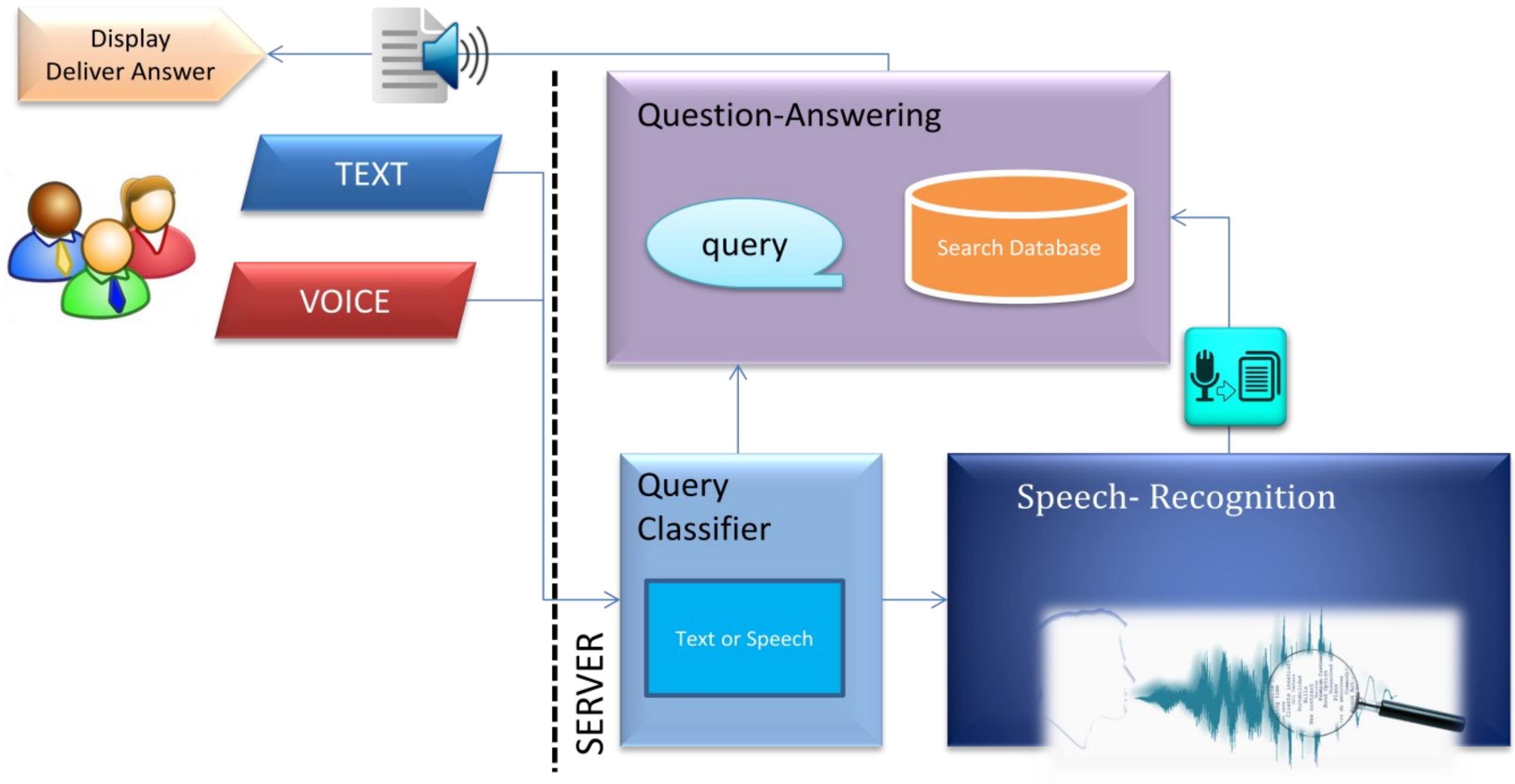

Figure 3: Personal Digital Assistant Architecture

Basically, at the higher level, it includes functions for voice recognition and text to speech conversion so that when an operator asks a question, it could be recognized and translated into a query to be processed at the bottom layer of the architecture where relevant knowledge resources are coded, indexed and stored. Query results are then transmitted to the higher level of the architecture where conversion into voice records and delivery occur. At this stage the operator can ask further or related questions. It is worth mentioning that the system is thought to be intelligent and intelligence is not only the ability to find out proper responses delivering meaningful contents but relies also on the capability to learn and evolve over time; thanks to advanced functions for human behavior tracing, the system is able to capture relevant features while interactions occur and provide proactive suggestions (accordingly). To this end, SOPHOS-MS relies on AI and machine learning algorithms that are part of its architecture (part of the question answering module).

## 4. OVERVIEW, REQUIREMENTS, DESIGN AND IMPLEMENTATION

As proof of concept the SOPHOS-MS system has been designed and implemented. It is based on the methodological framework described in Section 3 and its main requirements include:

- Provide plants operators with real time feedbacks and augmented reality contents on tasks/procedures execution so as to minimize the risk of accidents and support training.
- Provide plant operators with a personal digital knowledgeable assistant to interact with in order to gain data, information or knowledge about components, machines, tasks, procedures and processes. Interaction is in the form of vocal exchange according to a Q&A approach.

Moreover, in line with the Industry 4.0 principles other requirements include interoperability with legacy systems such as enterprise manufacturing systems, regulatory compliance, manageability and usability. Based on these requirements, the architecture depicted in Figure 2 has been designed.

From a conceptual point of view the architecture depicted in figure results in the following functionalities for its intended users:

- let operators be immersed and interact with a cyber space where he can gain meaningful insights, based on virtual and augmented reality or through a direct vocal exchange with a personal digital assistant, about man-machine interaction procedures to be compliant with safety standards and principles.

- exploit virtual and AR resources for operators' preliminary training on high risk tasks.

- support operators providing information that is usually not available in the workplace (i.e. machine productivity, expected maintenance operations, warning on unexpected dangers, risks that are likely to occur, suggestions on how to increase productivity, etc) as well as operator's training.

- send warning messages about the outcomes of improper operations (i.e. what happens if a maintenance operation is not performed, it the operators fails, etc).

To this end, the proposed architecture is scalable and modular in nature since the underpinning principle is the separation of the service infrastructure from the resources infrastructure. The service infrastructure includes a service manager and a service server that work together to:

- receive and process user's inputs from wearable and mobile devices,
- activate callback functions to retrieve proper resources from the underlying architecture,
- interact with user's devices for service delivery

On the other side, the Resource manager is responsible for contents (multimedia resources) and metadata processing and management (even through artificial intelligence) while a cloud infrastructure is devoted to storage and retrieval. In particular, the resource manager implements natural language processing methods that enable the personal digital knowledge navigator to interact properly with human operators as well as inference algorithms needed to acquire, manipulate and generate knowledge.

As for interfaces, the tool is envisaged for deployment at operational level and therefore end-users may not have advanced technical skills. Hence a user-centered design approach has been adopted and methods such as Hallway testing and Expert review based on Nielsen's usability heuristics have been applied.

For design and implementation purposes UML diagrams have been drawn up. In particular, structure diagrams such as the class diagram and the model diagram have supported the identification of the system components and relations at different levels of abstraction as shown in figure.

The class diagram (Figure 4) offers an overview of the system application domain where operators with specific roles and duties are called to interact with one or more machines executing one or more procedures; in doing so they can be supported by the SOPHOS-MS system that can be accessed through mobile devices and is able to provide both AR contents as well a personal digital assistant that allows vocal interactions as it would happen if the operator were involved in a conversation with a knowledgeable expert.

To this end, the SOPHOS-MS font-end application interacts with a Service Manager and a Service Server that activate a set of methods and functions needed to collect, retrieve and display contents. Moreover the Service Manager, in turn, interacts with the Service Server that implements AR algorithms and with the

Resource Manager that is responsible for contents (data, metadata, models, multimedia, knowledge etc) handling

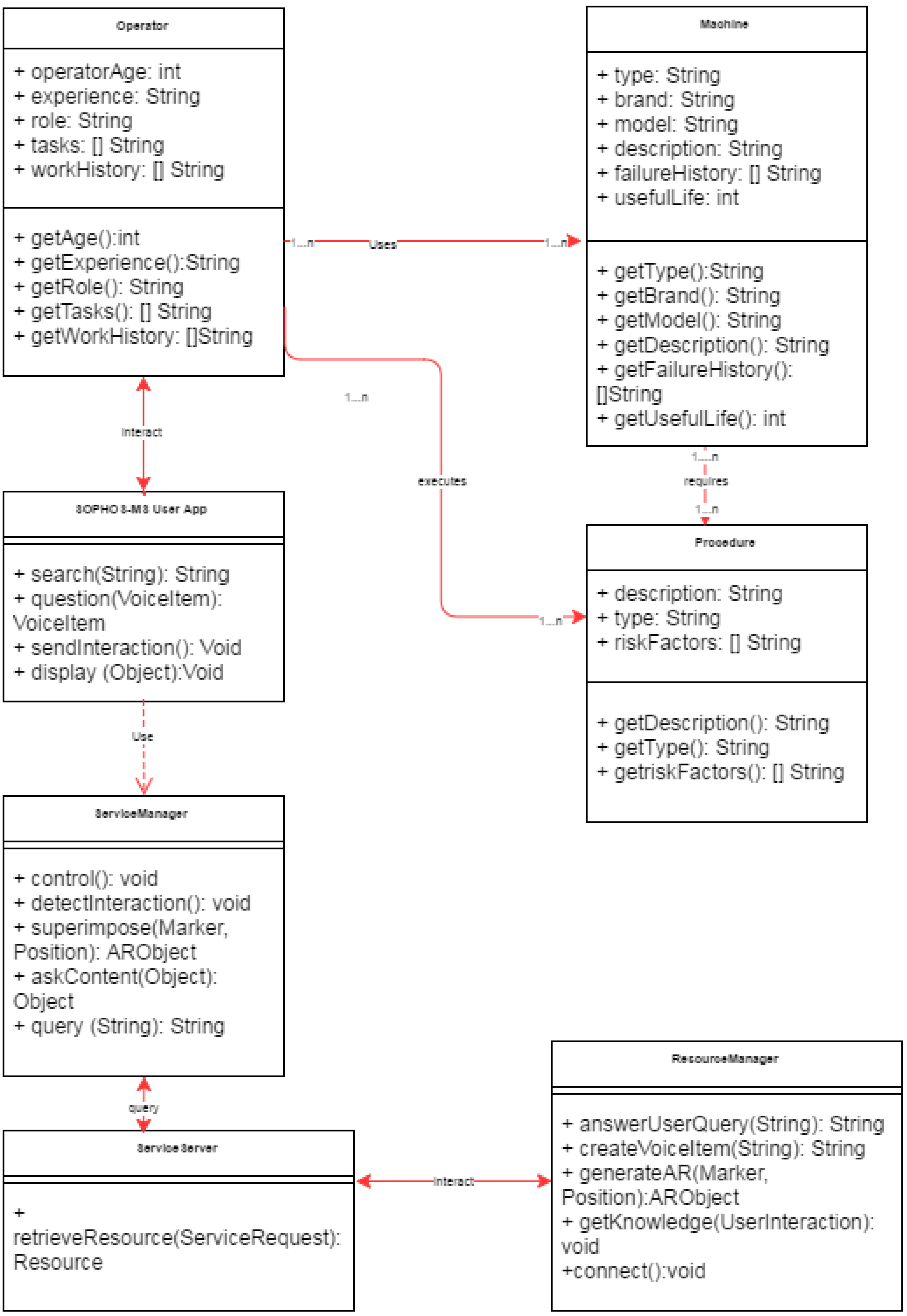

**Figure 4: Class Diagram**

A complementary view is provided in the model diagram (Figure 5). Here, the overall model has been broken down into three main layers. The presentation layer includes interfaces and applications on the user side and is devoted to ensure that the contents passing through are in the appropriate form for the recipient providing also interaction mechanisms (i.e. access to video, 3d representations of a particular machine, documents, etc). The application layer implements the core functionalities and the logic of the system in order to met both functional and non-functional requirements. The contents layer, instead, provides access to all those contents (information, multimedia, geometric models, metadata etc) that are needed to let the SOPHOS-MS system deliver meaningful information to its intended users.

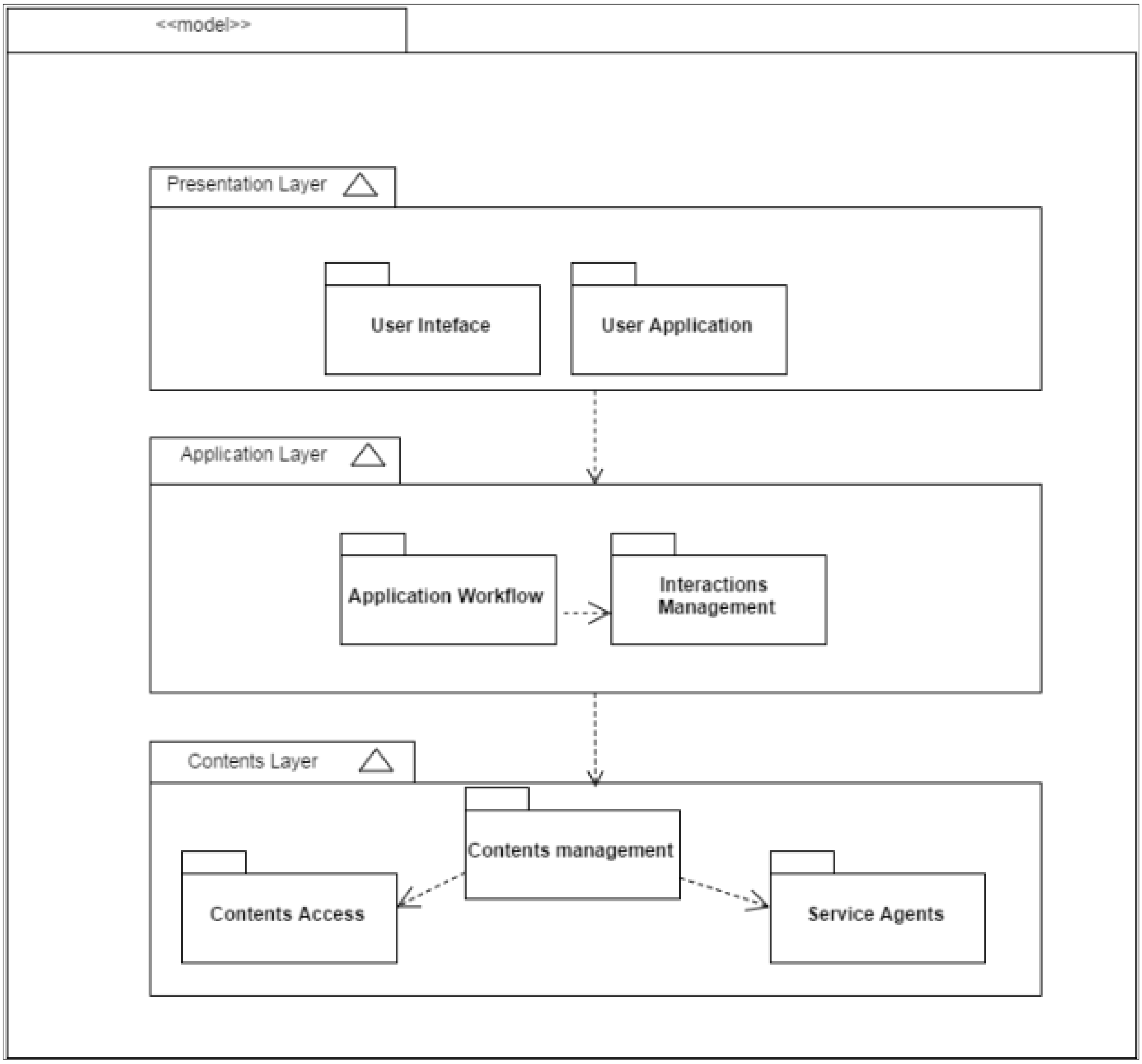

Figure 5: Model Diagram

On the other hand, the system behavior has been represented by the use case diagram (Figure 6), the state machine diagram and the sequence diagram. The use case diagram allows detecting the main requirements and functionalities of the system from the operator's perspective. In fact, the operator is supposed to access the SOPHOS-MS front-end application where he can search and/or visualize contents and communicate, in the form of a direct speech, with a personal digital assistant.

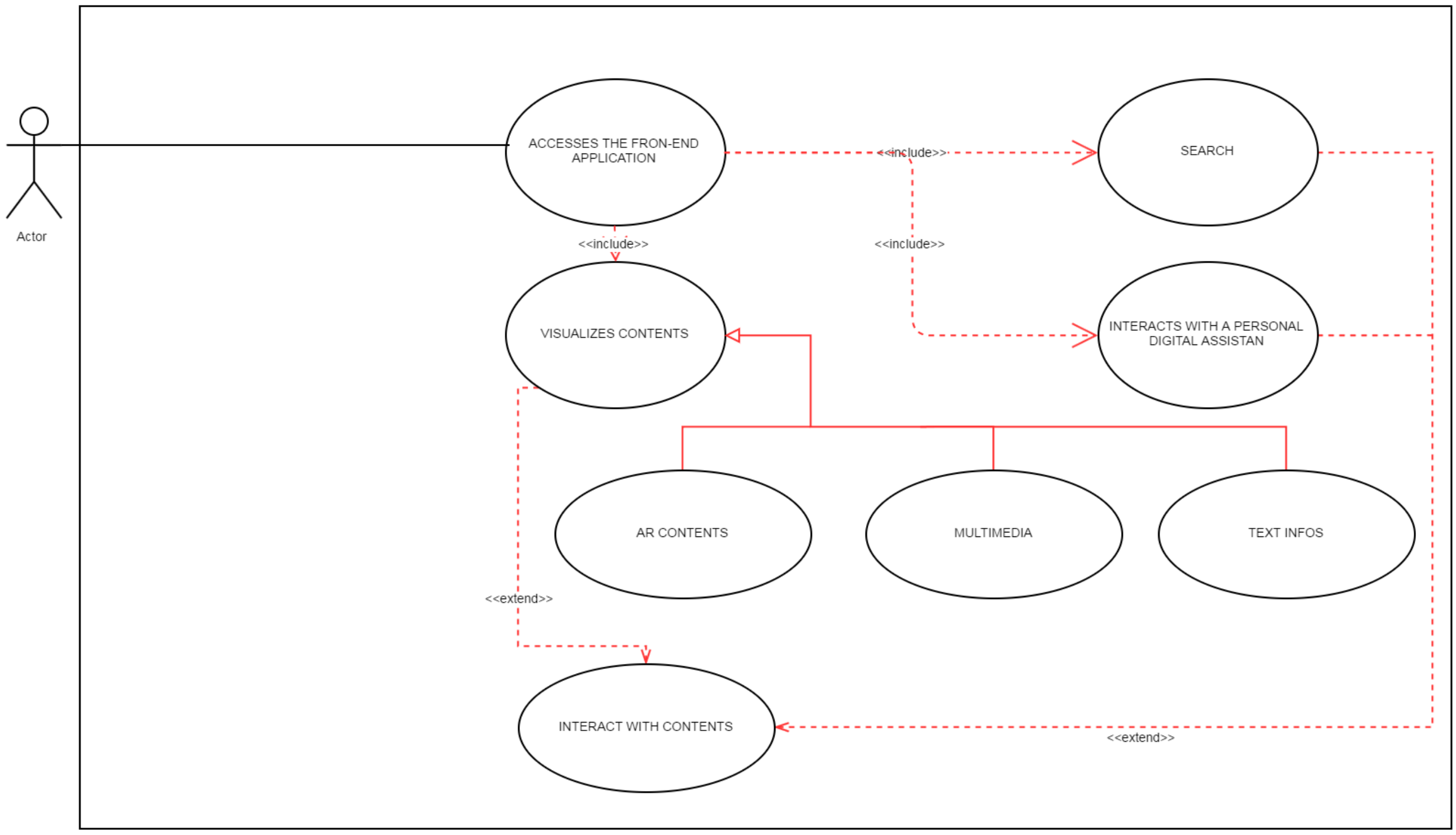

**Figure 6: Use Case Diagram**

The front end application is thought to be available on mobile and wearable smart devices so that, after contents are displayed, the operator can freely interact with them i.e. zoom in and zoom out of 3 representations, read text info, interact with virtual reality and/or augmented reality contents, etc. As of communication exchanges with the personal digital assistant, it can be accessed at any time and can end up with access to digital resources or manual search. However, interaction patterns are captured within the state machine diagram (Figure 7) where it is clearly highlighted that transitions are due to users' choices highlighting the event-driven nature of the SOPHOS-MS system. Finally, the SOPHOS-MS overall picture is completed by the sequence diagram where communication exchange are investigated and defined. Here, the attention is drawn on how system objects interact (Figure 8).

The conceptual design has been the basic prerequisite for the practical implementation of the proposed solution so that it could take shape and be actually deployed. To this end while many components have been developed from scratch coding functions and algorithms, other requirements/functionalities have been built upon already available suites. UNITY 3D has been selected as Integrated development Environment for the front-end application due to its great flexibility in terms of application deployment over different mobile and wearable devices as well as for interoperability purposes and interfacing with both existing systems (for information retrieval) and the underlying technical back-end architecture for real-time interaction management and contents provision. For augmented reality contents generation and delivery it has been necessary first to generate and encode specific markers that could be recognized from the camera of the device where the SOPHOS-MS is installed. Once the camera detects and recognizes a particular marker the associated virtual contents are superimposed on it and the operator is immersed in a digital space whereby he can fully interact. Markers generation and encoding has been carried out using the plugin Vuforia by Qualcomm TM supported by Unity 3D. Within the development environment each marker has been associated with one or more corresponding virtual 3D objects representing the total or partial reconstruction of real items/components/machines.

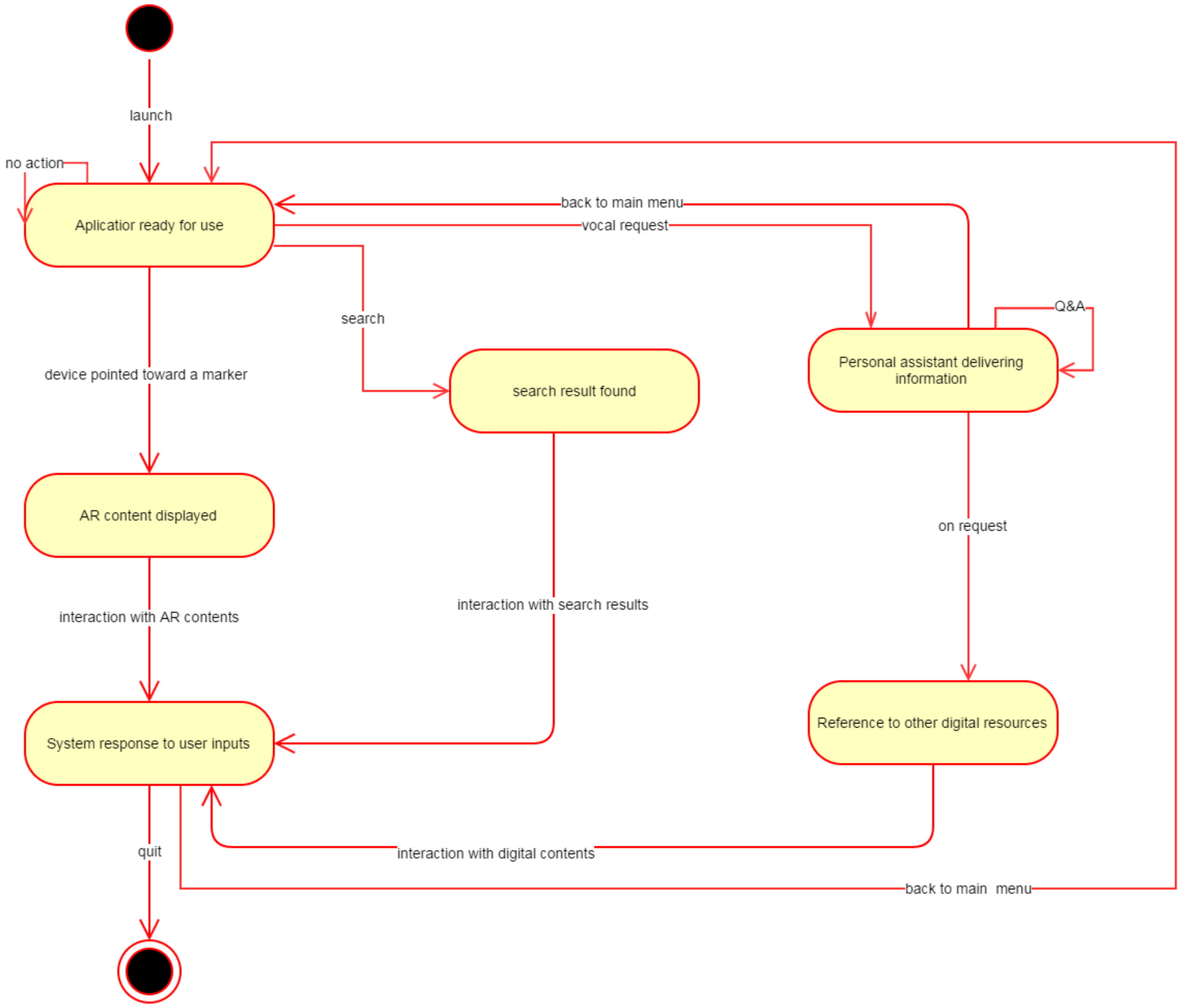

Figure 7: State Machine Diagram

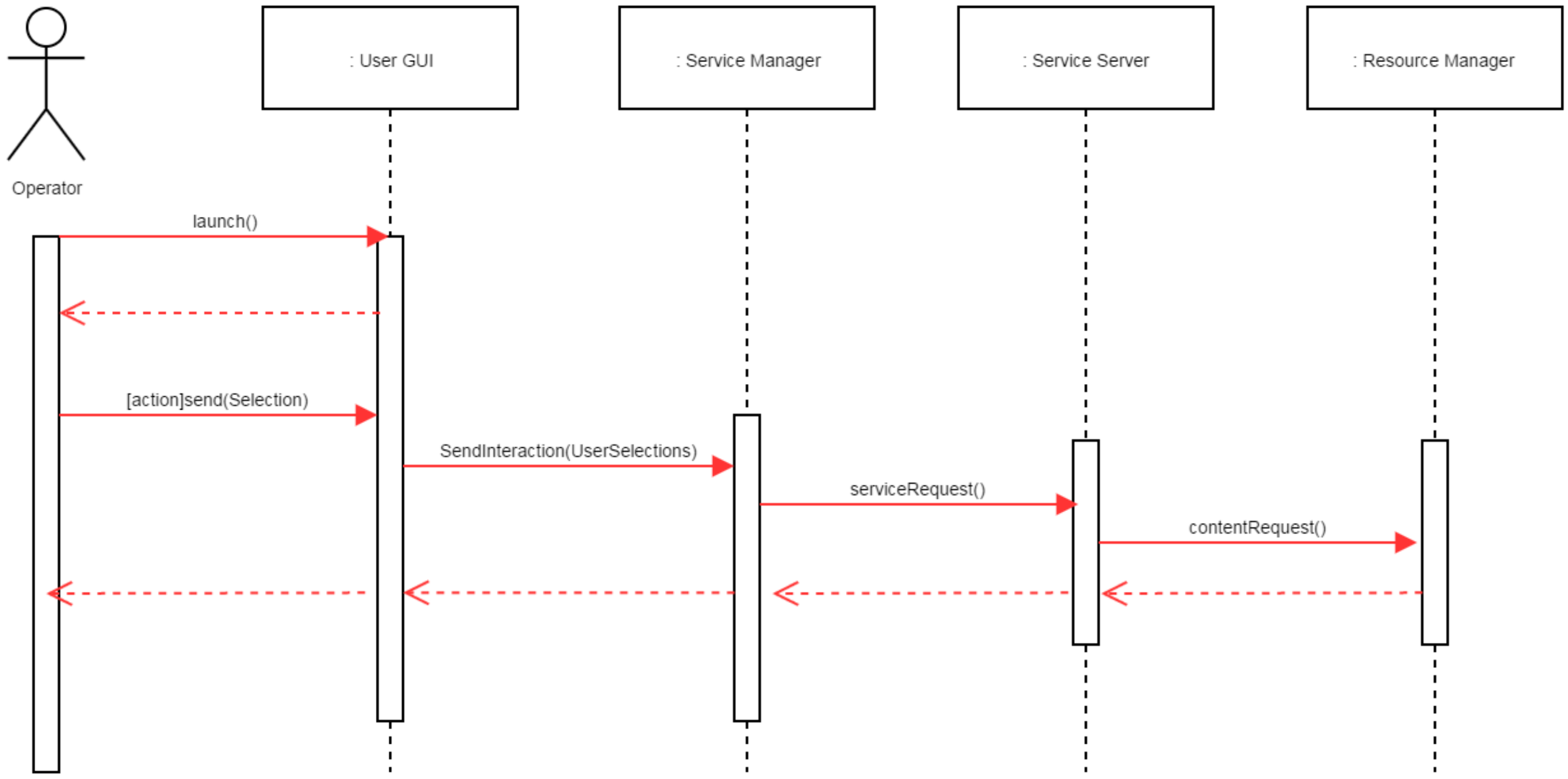

Figure 8: Sequence Diagram

Considering that a marker is a two-dimensional object it has been also necessary to rescale 3D virtual objects respect to the marker area and the third dimension. AR contents quality has been also widely investigated. It has been noticed that the main parameters called into question involve the marker complexity in terms of pattern and the quality of the camera used to capture the marker. As the camera resolution and the marker pattern increase the visual quality of the content enhances. Visualization, indeed, results from triangulation techniques applied between the marker points and the 3D object mesh so the best trade-off among marker complexity, camera performances and visual quality requirements has been searched. 3D Objects have been also developed through CAD tools seeking to combine the need for high graphic rendering with the requirement to avoid computational overload for users' devices.

Another key element of the proposed solution includes the virtual knowledge assistant acting as a knowledgeable expert. Ad previously discussed it includes: vocal recognition algorithms, functions for speech-to-text and text-to-speech conversion, research functions that are activated upon operator's interactions, contents indexing and proper matching and correlation algorithms to implement the capacity of providing proactive suggestions. The personal digital assistant is activated through the SOPHOS-MS graphical user interface. At this stage vocal interaction is handled by the voice manager subsystem that is responsible for the voice interaction management. It includes conversion from voice to text when the user asks a question and vice versa from text to speech when responses are gathered from the underling architecture. Once the user vocal request has been received and processed in text format it is transmitted the service manager that implements the functions needed to process the request and send it to the service server. In turn the service server interacts with the resource manager where a knowledge base as well as search functions have been implemented to retrieve the contents of interest. In particular search functions rely on the Lucene search engine that is an open source solution while their effectiveness is related to contents organization and indexing within the knowledge base. To this end, a particular effort has been done toward the definition of a proper contents infrastructure. Contents have been coded in XML format and based on a hierarchical tree structure able to grow and evolve over time in a fast and effective way without causing the need for redesign so that new contents can be added at any time without hindering the solution functioning. Contents have been also indexed through XML tags on key words to let search functions be faster and more effective. The system intelligence has also been implemented so that it is able to recognize those words that can be interchangeably used and when to many results matches are found it suggests the user to word the question differently. Moreover, thanks to the contents organization the system is given also with the intelligence to infer and detect related resources that could fall under the operator interests (acting in a proactive way) or could valuably support him in executing his tasks.

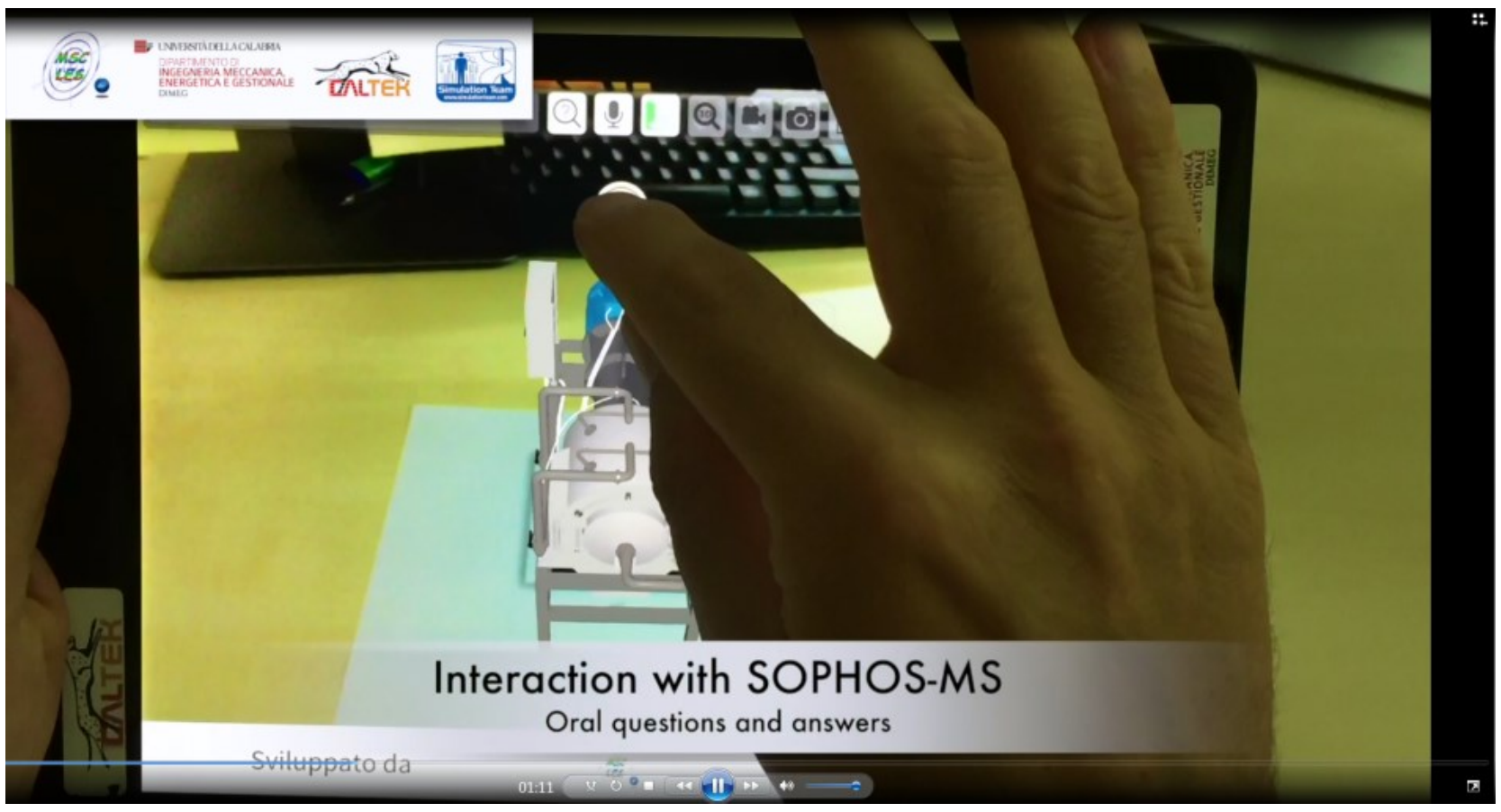

Figure 9: Oral questions and answers with SOPHOS-MS

## APPLICATION EXAMPLE

The solution deployment has been investigated through a particular use case concerning a CNC Milling Machine. Based on the technical infrastructure previously discussed, the machine 3D virtual model has been recreated and thanks to the connection with the enterprise system all relevant resources including text, images, videos have been imported.

This wealth of resources has been the starting point for developing ad hoc custom AR contents related to the machine operational model as well as the safety and maintenance procedures it requires. For instance Figure 10 shows the virtual representation of the machine. The application gives access also to 3D AR animations explaining safety and maintenances procedures required along the period of operation (Figure 11). Regular and preventive maintenance operations such as functional checks, corrective adjustments, tests for wear, parts exchange and repair have been carefully considered. Each procedure is broken down into steps and each step is visually shown through an augmented 3D virtual animation and vocally explained by the personal digital assistant. Along the explanation, the system is also able to send visual and vocal messages drawing the attention on potential dangers or risks the operator may incur. While accessing such resources the operator is also enabled to interact with the personal digital assistant asking explanatory questions in voice format. Beside the operator can even search for additional contents typing keywords in a text search box.

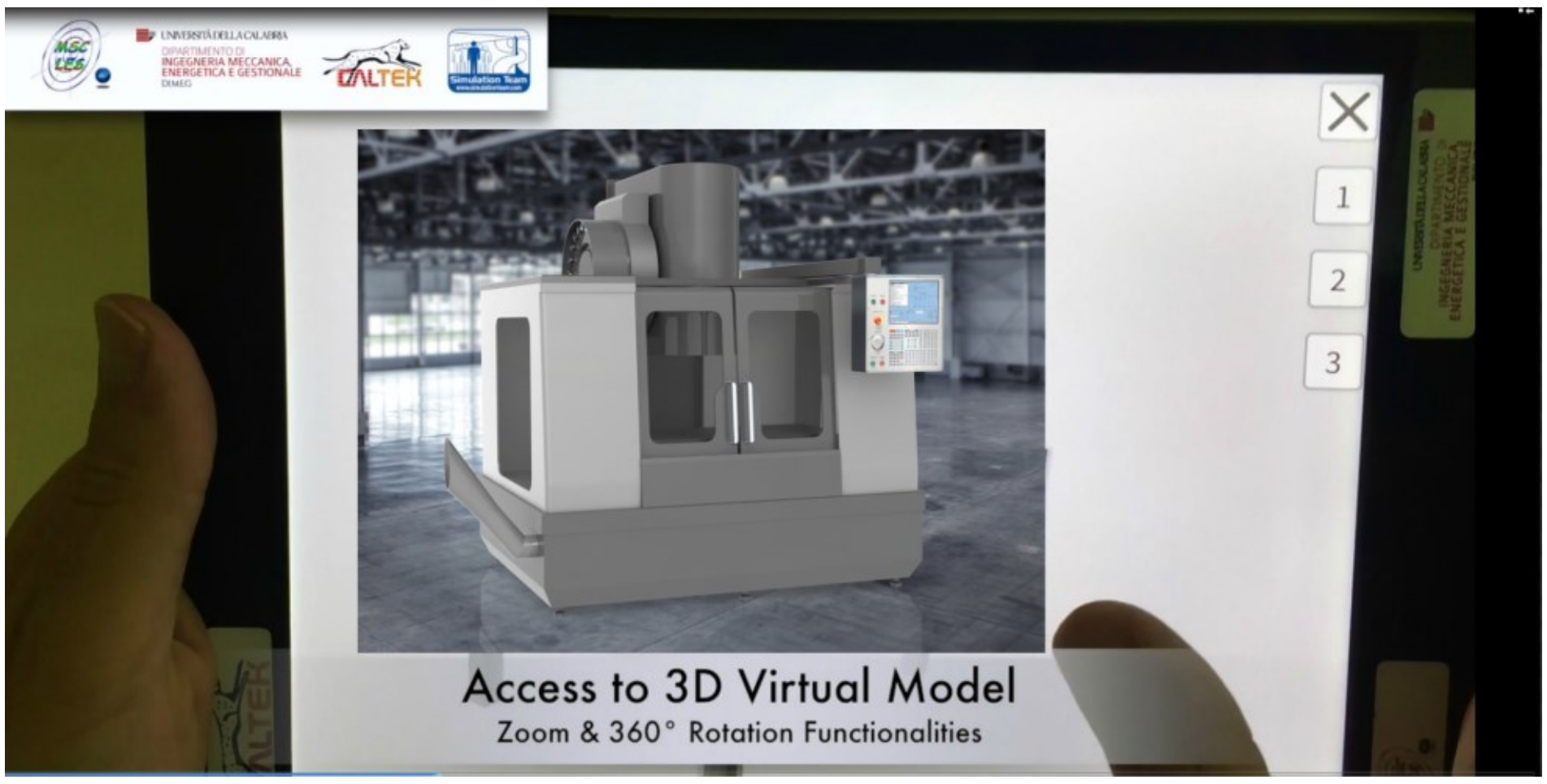

**Figure 10: Access to 3D Virtual Resources**

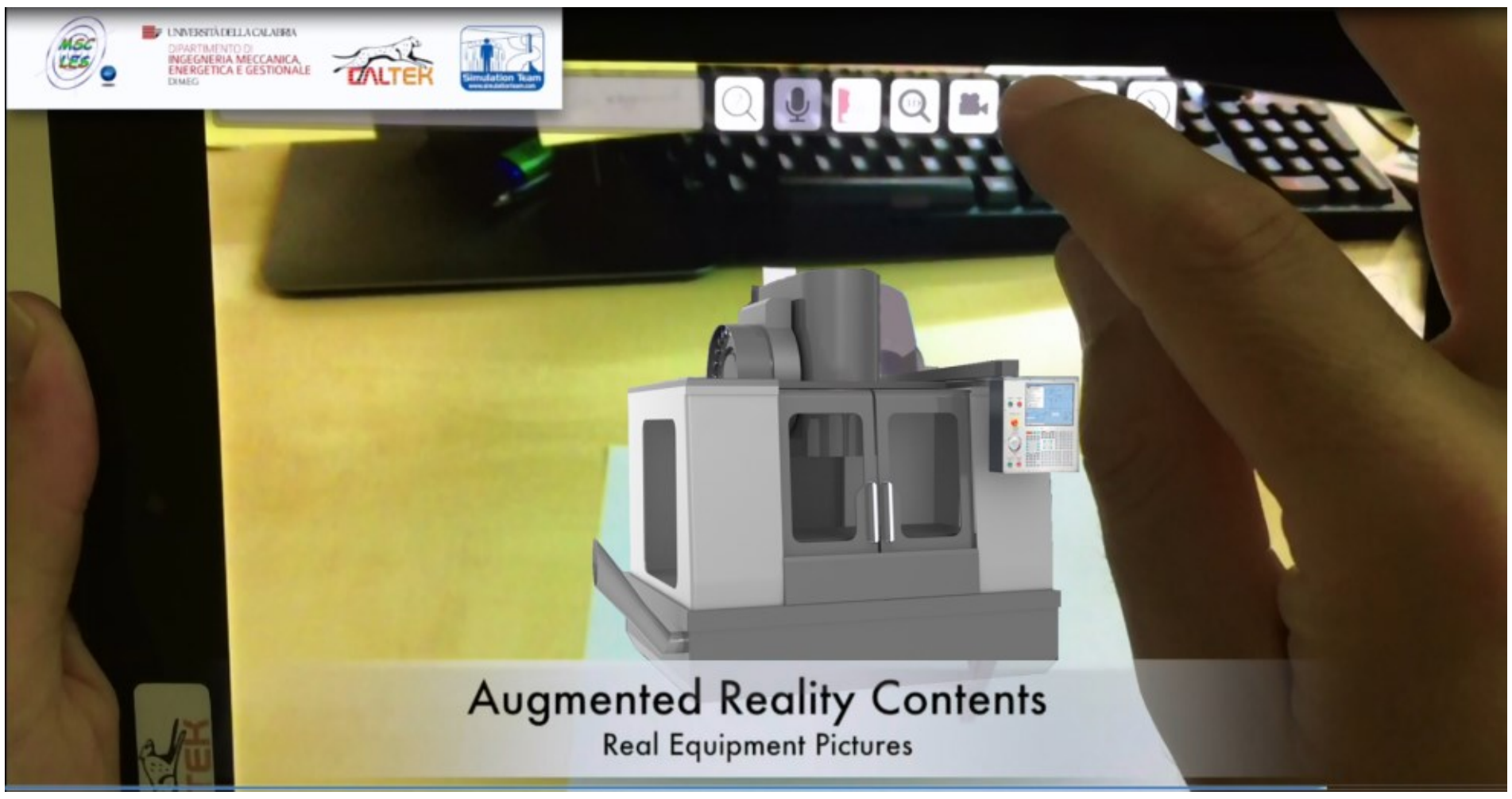

**Figure 11: SOPHOS-MS AR Contents**

In greater detail, the SOPHOS-MS application can support and train operators on operational procedures involving the machine usage as well as maintenance operation involving several inspection points such as safety devices, pneumatic/hydraulic systems, filters, axis, electrical system, spindle/transmission, coolant system, operator panel, accessories, lubrication system, tool changing system, turret, etc. Hence SOPHOS-MS is devised to deliver information on:

- safety measures to be taken along the machine period of operation (i.e. safety equipment to wear, ensure that machine guards are in position, before cleaning swarf accumulations switch off and bring the machine to a complete standstill,)
- potential hazards: hair/clothing entanglement, eye injury, skin irritation, metal splinters and burrs, flying debris, etc.
- start and stop the machine in normal and emergency situation as well as machine restart after emergency;
- milling operations, machine modes ( hand and automatic), equipment and control panel;
- messages that can be displayed and procedures to be activated upon specific messages are delivered;
- cutting fluids to be applied with different materials;
- quality control procedures and inspections checks;
- tools re-sharpening/replacing;

As for maintenance operations the system can deliver detailed instructions on how to lubricate way covers, clean the chips out, take the filter off and clean it, grease the working parts moving and grooving, check and restore at the right operating level:

- hydraulic pressure;
- fluids and lube;
- chuck pressure;
- cooling unit.

This way the SOPHOS-MS application is a tool that can be deployed both real-time and off-line for a variety of purposes such as :

- make available knowledge resources that usually may not be directly available in the workplace
- be consultation tool for operators to get easy and fast access to information and contents through advance fruition modes;
- an immersive and absorbing environment for preliminary training on new and /or complex procedures
- a tool for safety and security enhancement

To this end the SOPHOS-MS solution has been made compliant with different fruition technologies including tablets and mobile devices but not only. As a matter of facts, a particularly suitable solution for training can be obtained through headset (i.e the Samsung Gear VR) for contents display and a gesture control system like Myo Armbands for interaction with virtual reality (Figure 12 and Figure 13). In this configuration, SOPHOS-MS can be profitably used as an immersive training system that allows operators to get acquainted with new procedures or safety measures in a safe environment.

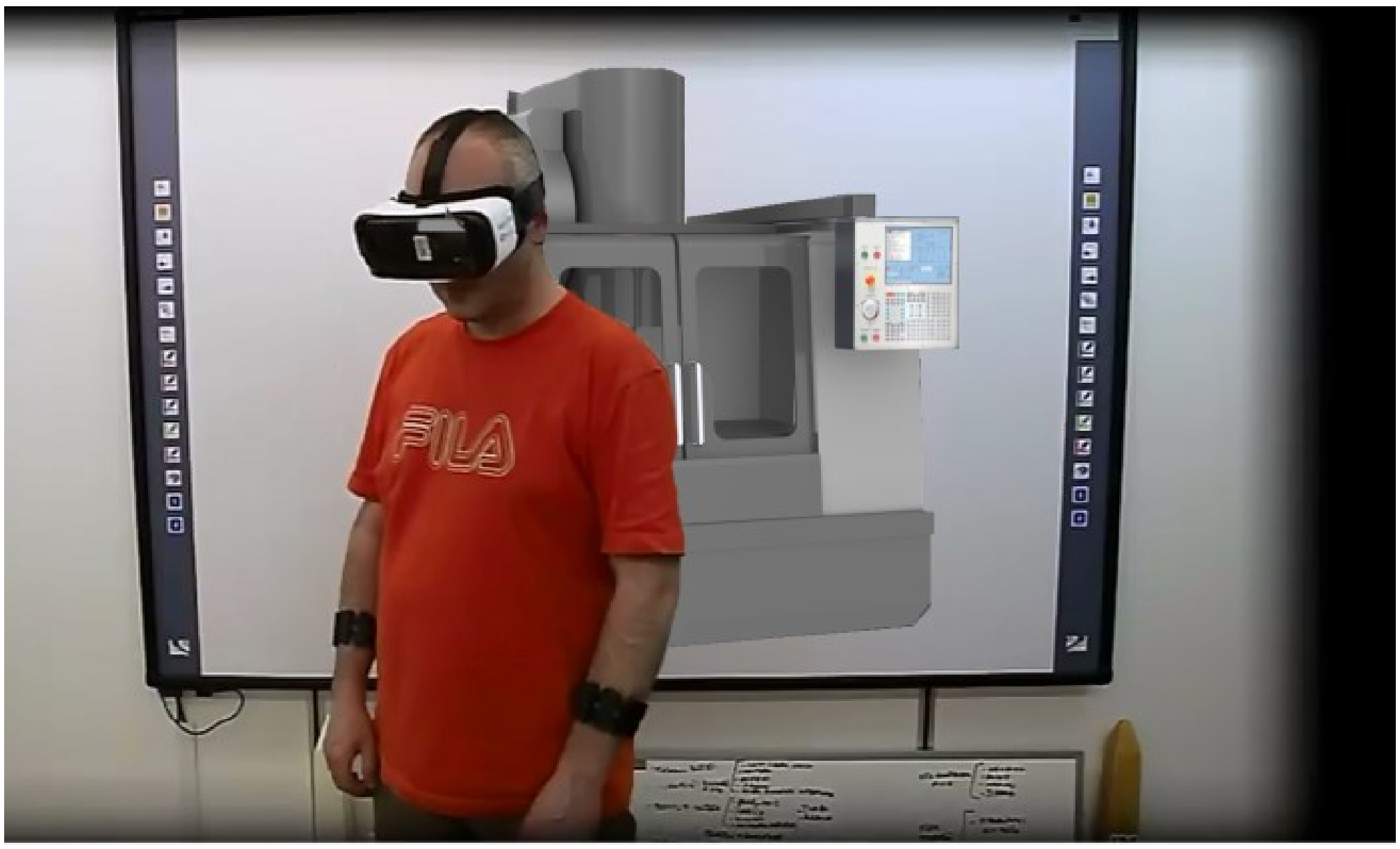

Figure 12: SOPHOS-MS integrated with immersive headset, interactive witheboard and motion capture armbands

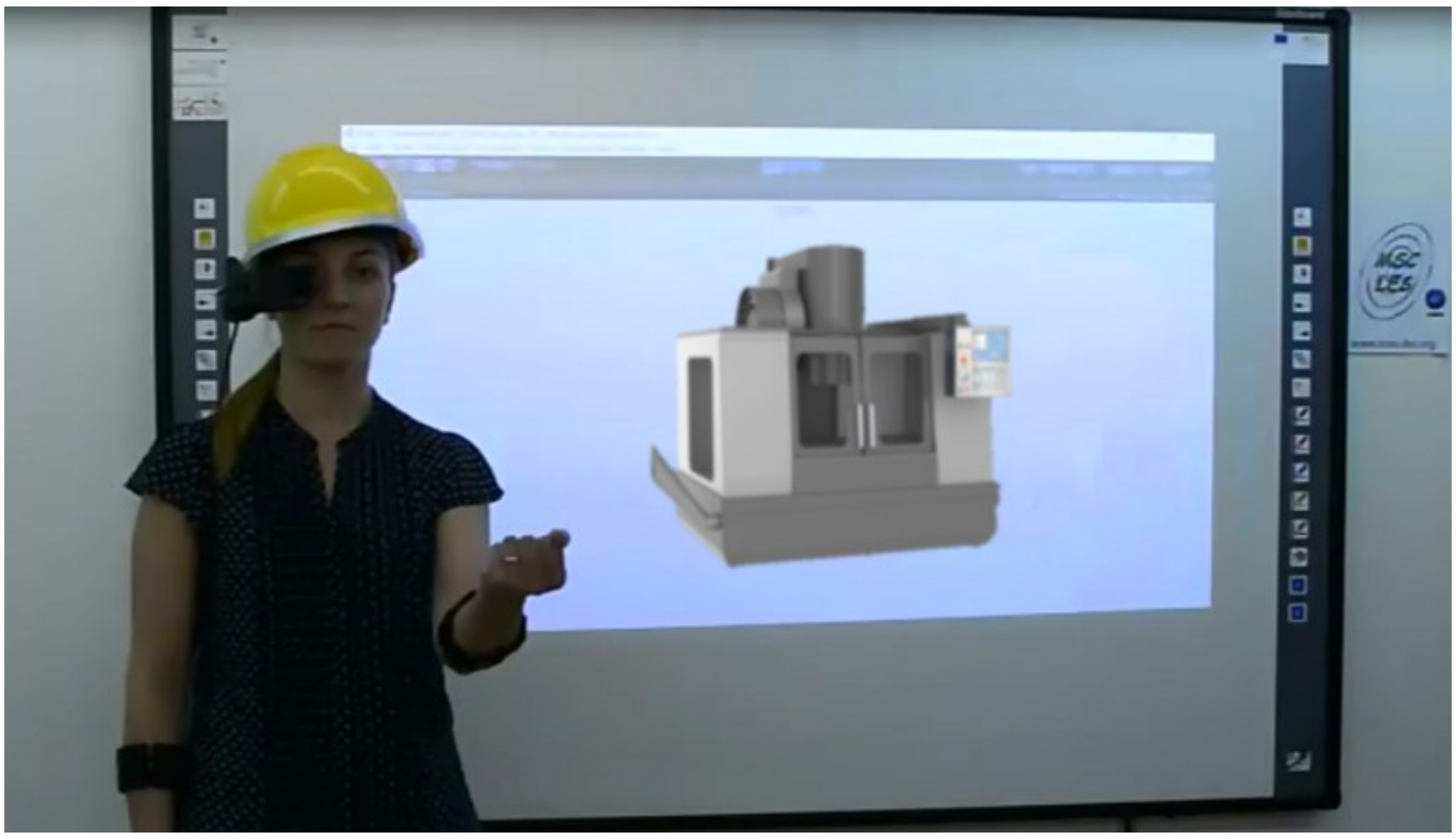

Figure 13: SOPHOS-MS Integration with monocular smart glasses, motion capture armbands, interactive witheboard

In this direction, a specific analysis has been carried out to assess how the proposed solution may impact on labor performance. The evaluation has been based on the comparison of traditionally trained operators with operators trained by SOPHOS-MS in job specific training. Given that the more effective the training is, the faster the learning rate is, empirical data have been collected considering two sets of operators with the same professional starting level. In particular, to ensure an unbiased analysis a sample of 20 totally inexperienced people with similar past working experiences has been taken. Training activities where focused on the CNC milling machine setup operations to be performed for cushion slides (Figure 14) manufacturing whose production data are reported in Table 1. The pull of available resources has been split into two subsets: 10 operators underwent traditional training classes while 10 operators were trained by SOPHOS-MS. Then each operator's performance has been directly observed and measured over a two weeks period or 64 batches production time taking into account how the average setup time changes for each set of trained operators as the cumulative production increases.

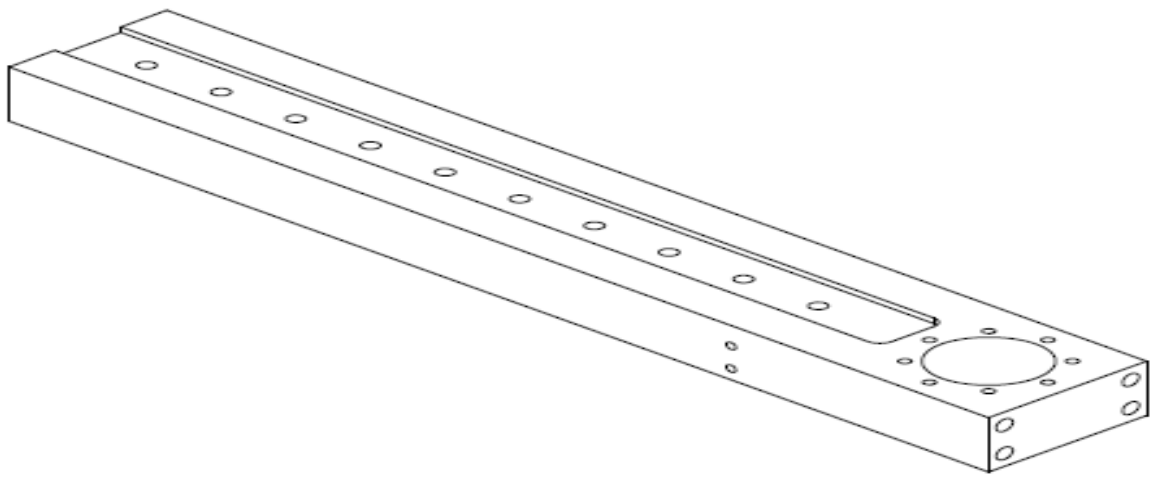

Figure 14: Manufactured part: cushion slide

This component is produced in batch of 100 pieces starting from steel bars low carbon 1018.

| Average cycle time | Cycle time variability | Average setup time | Setup time variability | Defect rate |
|---|---|---|---|---|
| 01:23:14 | 10.63% | 16.97 minutes | 12.33% | 2% |

Table 1: Process parameters

The learning curve model that has been referred to is the Towill and Cherrington (1994) whose expression in given in (1).

$$Y_i = B_0 + B_1 x^{-B_2} \quad (1)$$

The main learning curve parameters include:

$Y_i$ the cumulative average setup time given by the total setup time over one week (5 working days) divided by the number of batches produced within the same time frame.

$B_0$ : the asymptote that is the minimum achievable setup time,

$B_1$ :the maximum possible reduction (i.e. the difference between the setup time of the first batch and $B_0$), $x$ : the total number of batches produced,

$B_2$ : the rate of change for each successive batch setup time as it moves toward the lower bound.

Results and data are reported in tables 2a and 2b as well as in figures 15 and 16. The analysis of available data highlights that (as expected) a learning effect exists and, as a result, in both cases (traditional and SOPHOS-MS based training) the average cumulative setup time decreases with the number of manufactured components (as shown in figures 15 and 16). Furthermore the learning effect follows a predictable pattern: when production doubles the time required for manual operation decreases on average by 8.15% in case of traditionally trained operators and by 10.18% in case of operators trained by SOPHOS-MS. This result becomes even more meaningful considering that typical learning slopes for machining operations are in the range of 90%-95% which the average learning rate achieved through traditional training (91.85% as shown in table 2a) is fully compliant with. On the other hand, a greatest gain is achieved thanks to the SOPHOS-MS deployment (89.82% as shown in table 2b); here, the positive effect of the proposed solution can be explained considering that the learning effect is amplified by engaging as well as immersive technologies/tools that make training more effective.

| Cumulative Production | Average Unit Setup Time (min) | Parameter $B_2$ | Learning Rate R |
|---|---|---|---|
| 1 | 27.88 | | |
| 2 | 27.05 | -4.84871 | 97.0% |
| 4 | 25.72 | -2.38811 | 95.1% |
| 8 | 23.69 | -1.5525 | 92.1% |
| 16 | 21.39 | -1.12757 | 90.3% |
| 32 | 18.97 | -0.86746 | 88.7% |
| 64 | 16.68 | -0.69187 | 87.9% |

**Estimated learning rate: 91.85%**

Table 2a: Traditional training

| Cumulative Production | Average Unit Setup Time (min) | Parameter $B_2$ | Learning Rate R |
|---|---|---|---|
| 1 | 26.47 | | |
| 2 | 24.72 | -4.87874 | 93.4% |
| 4 | 22.62 | -2.37529 | 91.5% |
| 8 | 20.45 | -1.53499 | 90.4% |
| 16 | 18.18 | -1.10881 | 88.9% |
| 32 | 15.92 | -0.84885 | 87.6% |
| *64* | 13.87 | -0.67416 | 87.1% |

**Estimated learning rate: 89.82%**

Table 2b: Training with SOPHOS-MS

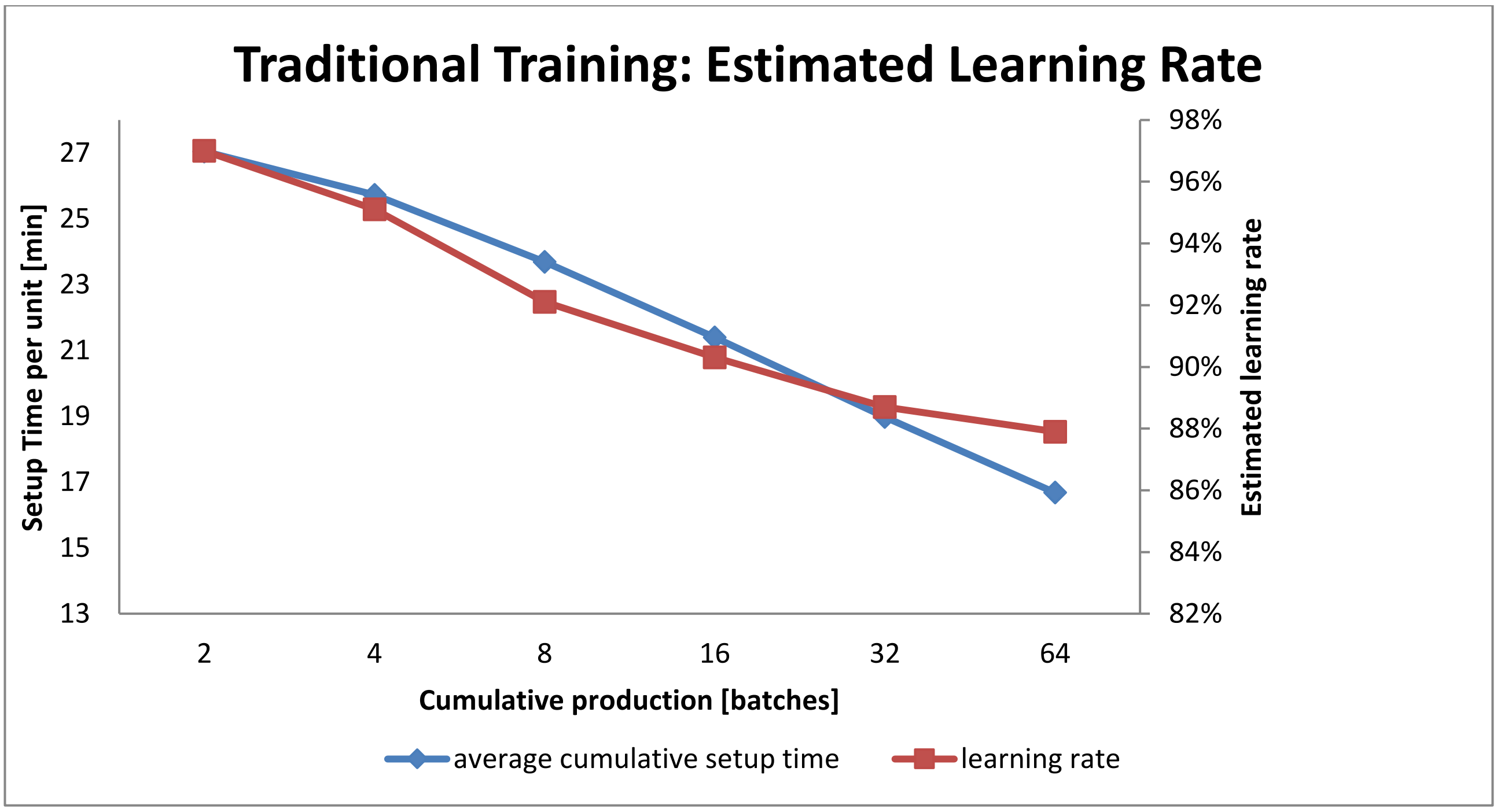

Figure 15: Estimated learning rate for traditional training

Furthermore as shown in figures 17 and 18 operators trained by SOPHOS-MS have, on average, better performances compared to traditionally trained operators: their initial productivity level is higher at their start and they keep on outperforming traditionally trained operators all along (recalling that a learning rate of 93.40% is better than 97.00%).

Indeed, the average marginal difference between the two sets of operators grows asymptotically along the learning curves as operators gain experience (see figure 17) demonstrating that SOPHOS-MS could be profitably used for initial training and make the difference between who uses it and who does not.

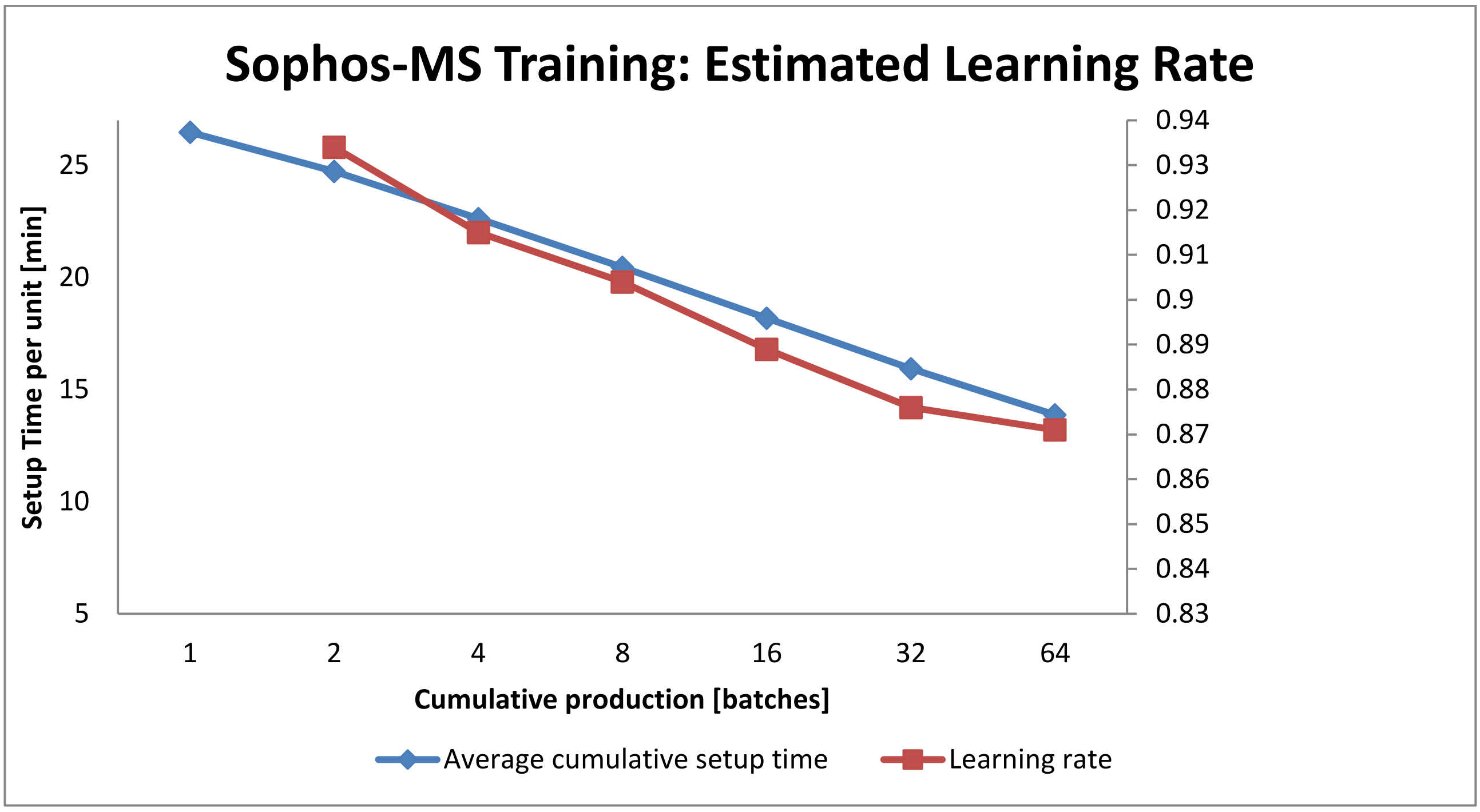

Figure 16: Estimated learning rate for SOPHOS-MS training

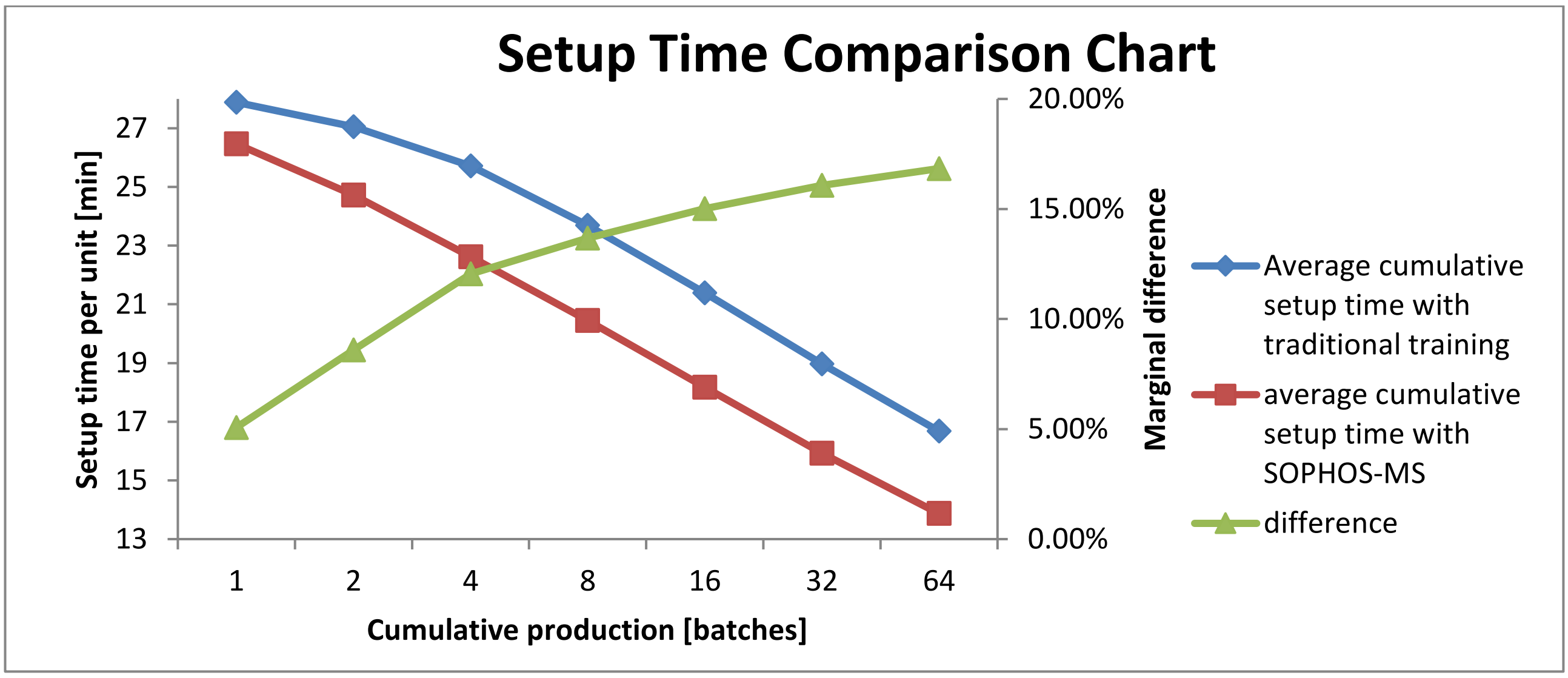

Figure 17: Comparison between average setup times achieved by traditional and SOPHOS-MS training

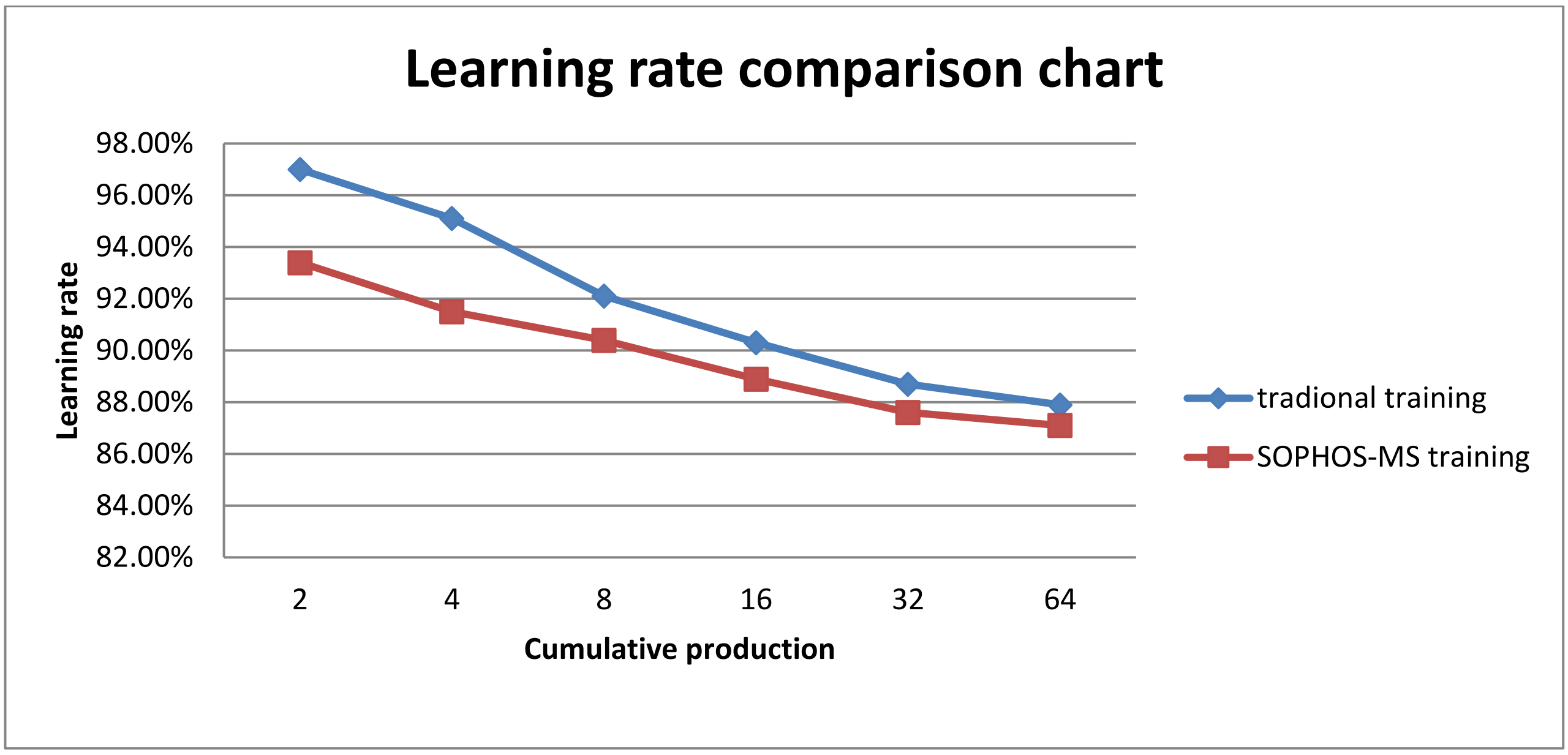

**Figure 18: Learning rates comparison**

It is worth pointing out that before comparing performances levels achieved with and without the proposed solution, statistical significance tests have been carried out in order to ascertain that the datasets related to traditionally trained operators and operators trained by SOPHOS-MS were significantly different. In other words significance tests have been carried out to assess whether observed differences between the two groups' averages reflect real differences or, conversely, sampling errors.

To this end the Mann-Whitney U test has been applied to assess whether the average unit setup times achieved by the two groups of operators for each target production level (1, 2,4,8,16,32, 64 batches) were statistically different from each other. As well known from statistics, it is a nonparametric but robust test that allows two groups to be compared without making the assumption that values are normally distributed. Besides, in the situation considered herein the assumptions that the Mann-Whitney U test requires are fulfilled, i.e. samples and observations are independent and observations are continuous and can be arranged in ascending order with no need to be normally distributed.

As previously mentioned the Mann-Whitney U test has been applied to the groups of observations collected for each target production level, namely the first batch production, 2nd batches production, 4th batch production, 8th batch production, 16th batch production, 32nd batch production and 64th batch production.

The test results for each production level that has been considered are reported in Figures 19-25 and in Tables 3-9.

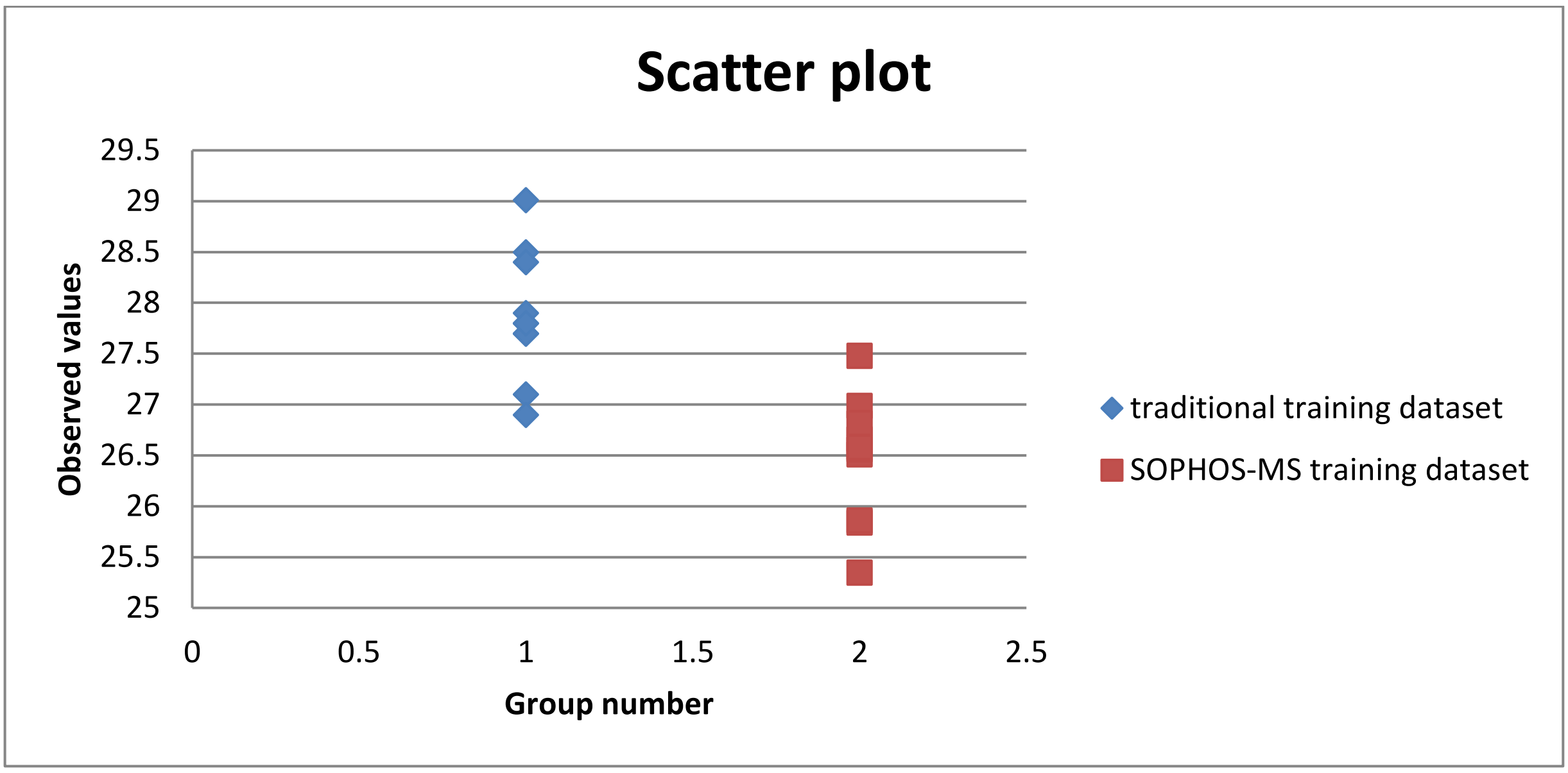

Figure 19: Scatter plot: setup times for the 1st batch production

| Sum of ranks for Group 1 | 152 | U statistic value | Critical value of U | Z-score | p-value |
|---|---|---|---|---|---|
| Sum of ranks for Group 2 | 58 | | | | |
| | | 3 | 23 | 3.552866 | 0.000381 |
| **Mann-Whitney U test** | | Reject the null hypothesis at the 0.05 significance level. Two populations are distributed differently. | | | |

Table 3: Significance test for average setup times in the first batch production

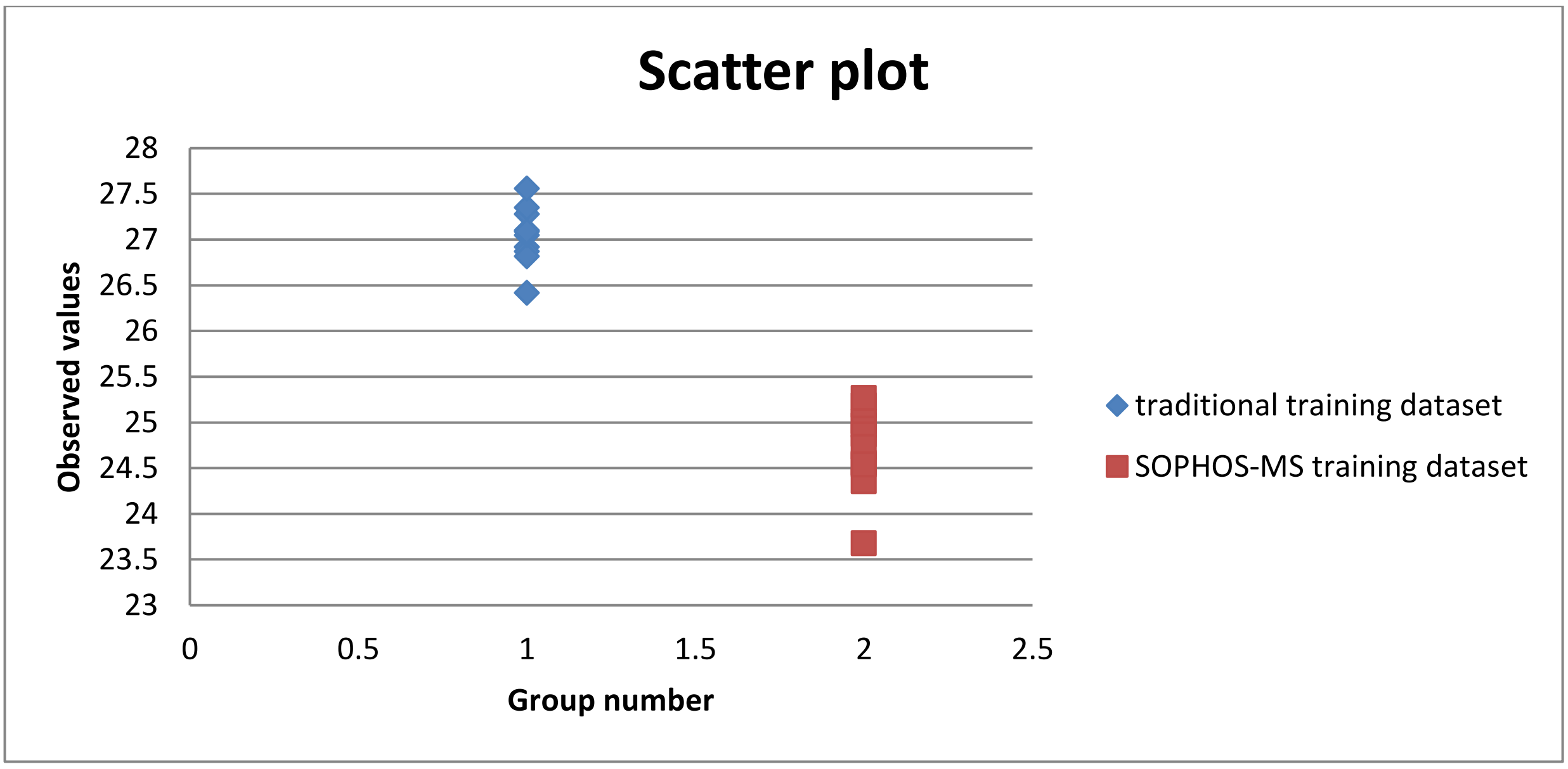

Figure 20: setup time for the second batch production

| Sum of ranks for Group 1 | 155 | U statistic value | Critical value of U | Z-score | p-value |
|---|---|---|---|---|---|
| Sum of ranks for Group 2 | 55 | | | | |

| | | | | |
|---|---|---|---|---|
| | 0 | 23 | 3.779645 | 0.000157 |
| **Mann-Whitney U test** | Reject the null hypothesis at the 0.05 significance level. Two populations are distributed differently. | | | |

Table 4: Significance test for setup times related to the second batch production

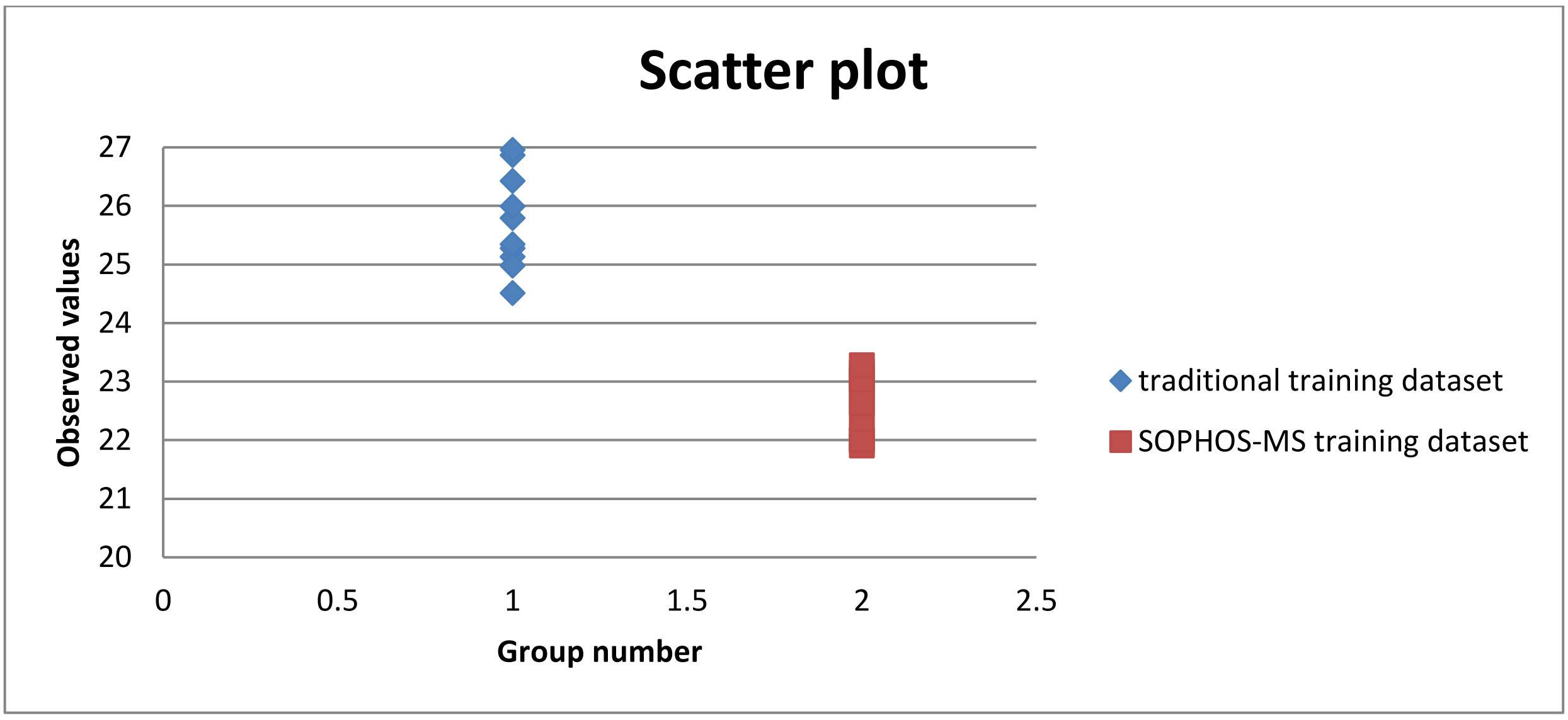

Figure 21:setup time for the 4th batch production

| | | | | | |
|---|---|---|---|---|---|
| **Sum of ranks for Group 1** | 155 | **U statistic value** | **Critical value of U** | **Z-score** | **p-value** |
| **Sum of ranks for Group 2** | 55 | | | | |
| | | 0 | 23 | 3.779645 | 0.000157 |
| **Mann-Whitney U test** | | Reject the null hypothesis at the 0.05 significance level. Two populations are distributed differently. | | | |

Table 5: Significance test for the setup time of the 4th batch production

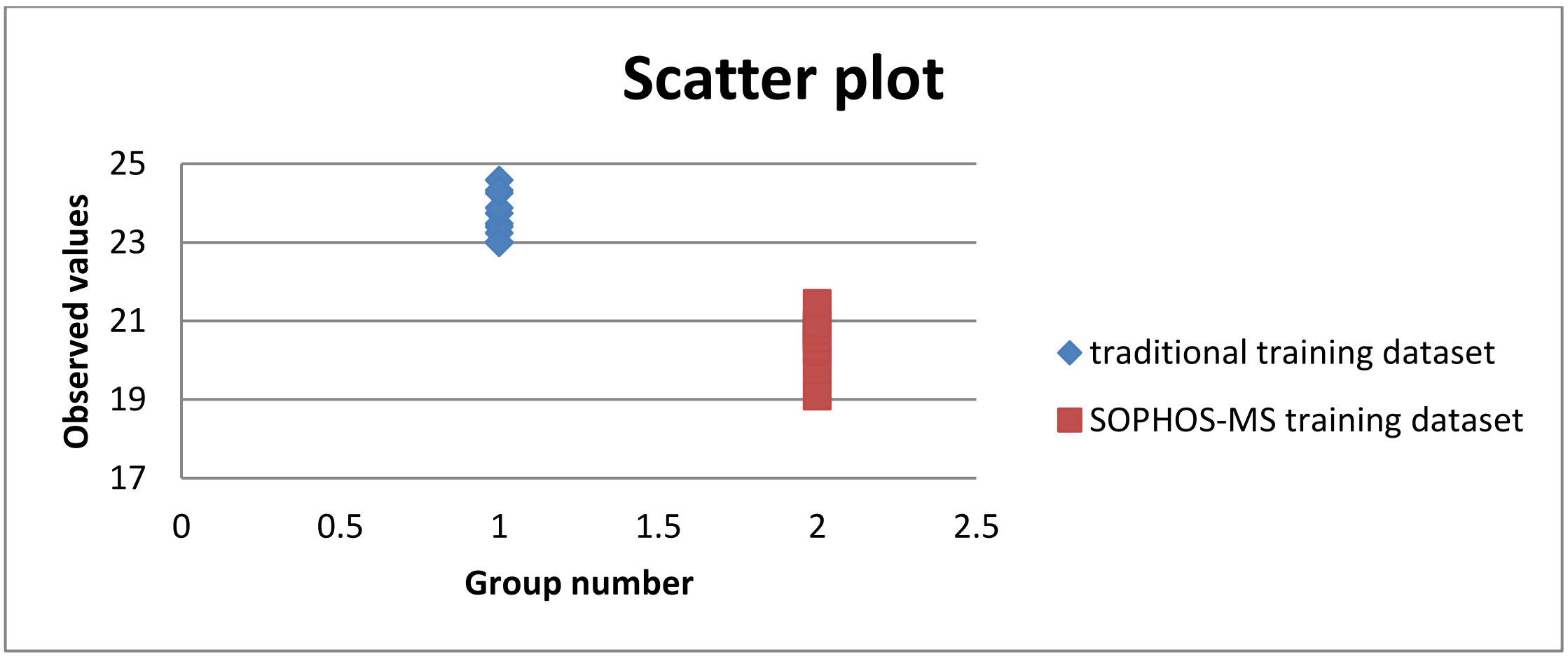

Figure 22: Figure 3:setup time for the 8th batch production

| Sum of ranks for Group 1 | 155 | U statistic value | Critical value of U | Z-score | p-value |
|---|---|---|---|---|---|
| Sum of ranks for Group 2 | 55 | | | | |
| | | 0 | 23 | 3.779645 | 0.000157 |
| **Mann-Whitney U test** | | Reject the null hypothesis at the 0.05 significance level. Two populations are distributed differently. | | | |

Table 6: Significance test for the setup time of the 8th batch production

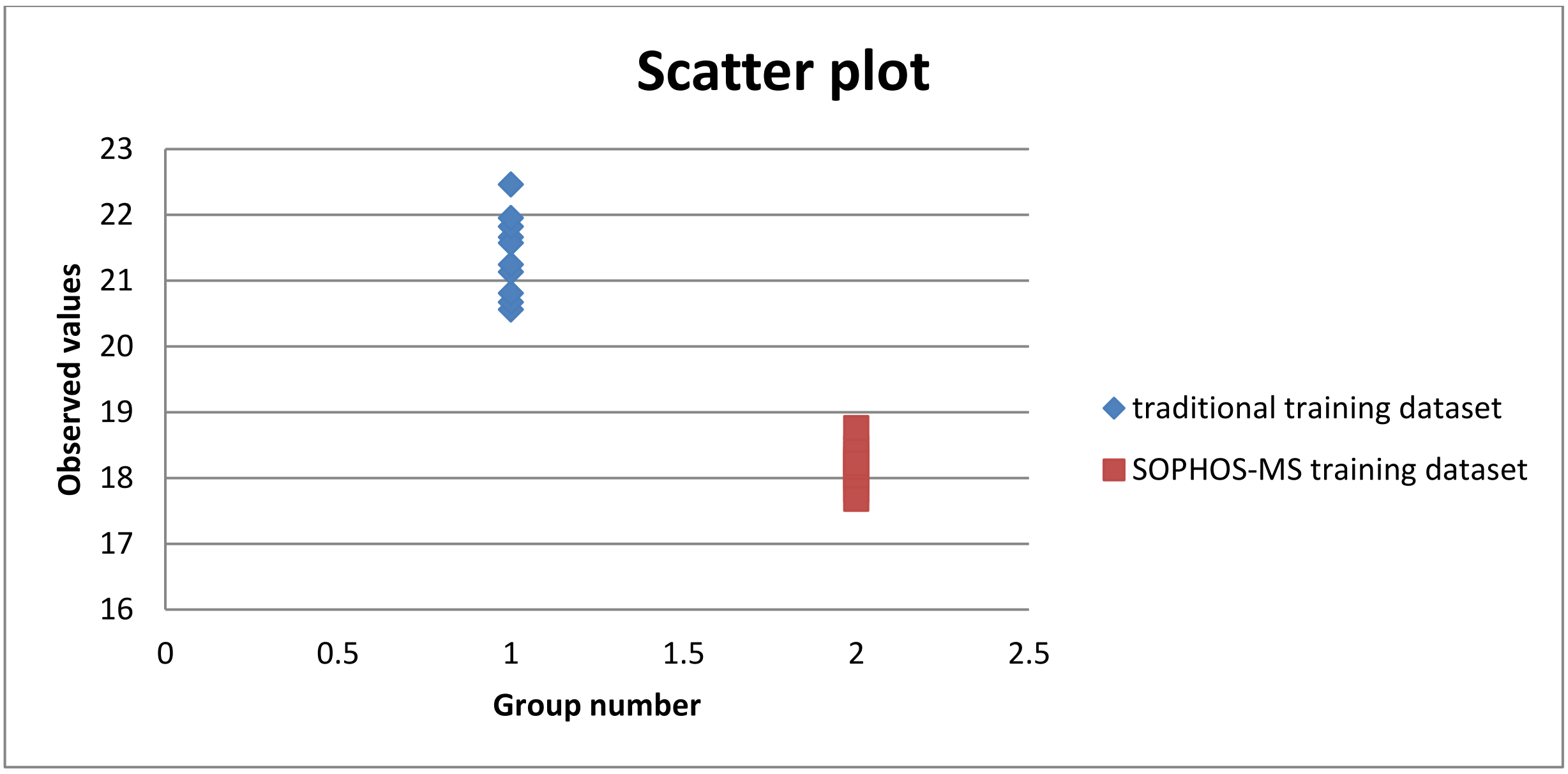

Figure 23: setup time for the 16th batch production

| Sum of ranks for Group 1 | 155 | U statistic value | Critical value of U | Z-score | p-value |
|---|---|---|---|---|---|
| Sum of ranks for Group 2 | 55 | | | | |
| | | 0 | 23 | 3.779645 | 0.000157 |
| **Mann-Whitney U test** | | Reject the null hypothesis at the 0.05 significance level. Two populations are distributed differently. | | | |

Table 7: Significance test for the setup time of the 16th batch production

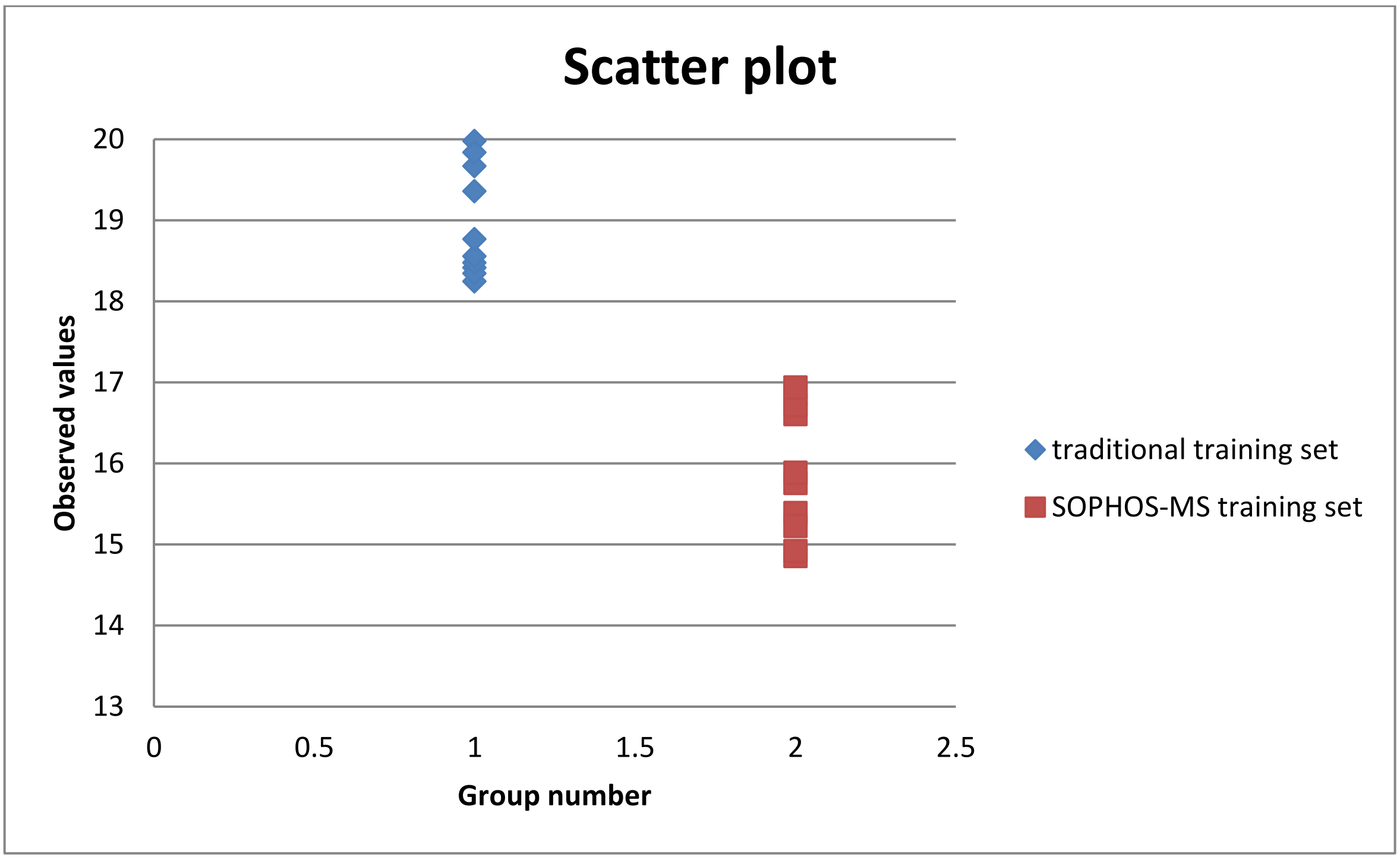

Figure 24: setup time for the 32nd batch production

| | | | | | |
|---|---|---|---|---|---|
| **Sum of ranks for Group 1** | 155 | **U statistic value** | **Critical value of U** | **Z-score** | **p-value** |
| **Sum of ranks for Group 2** | 55 | | | | |
| | | 0 | 23 | 3.779645 | 0.000157 |
| **Mann-Whitney U test** | | Reject the null hypothesis at the 0.05 significance level. Two populations are distributed differently. | | | |

Table 8: Significance test for the setup time of the 32 batch production

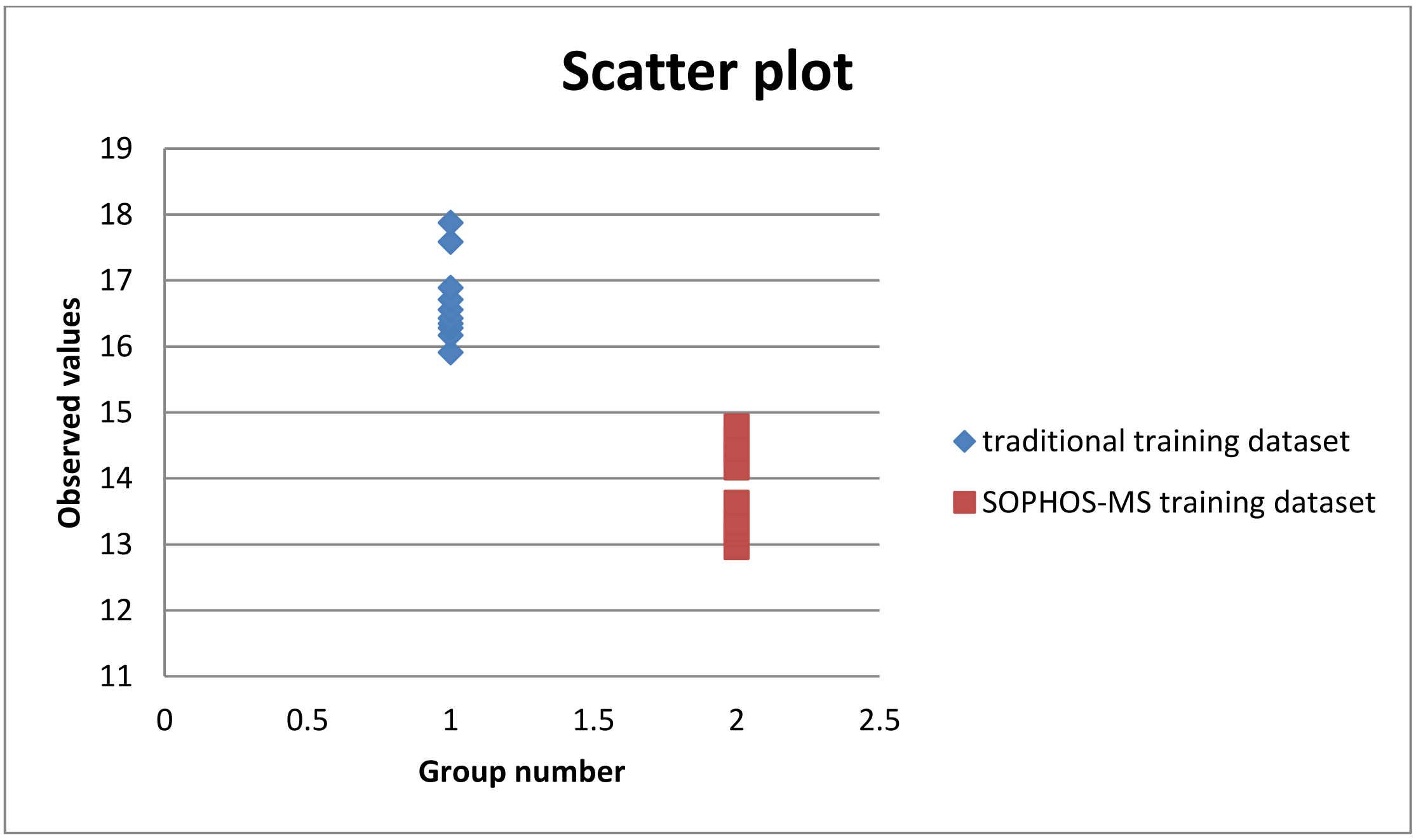

Figure 25: setup time for the 64th batch production

| | | | | | |
|---|---|---|---|---|---|
| **Sum of ranks for Group 1** | 155 | **U statistic value** | **Critical value of U** | **Z-score** | **p-value** |
| **Sum of ranks for Group 2** | 55 | | | | |
| | | 0 | 23 | 3.779645 | 0.000157 |
| **Mann-Whitney U test** | | Reject the null hypothesis at the 0.05 significance level. Two populations are distributed differently. | | | |

**Table 9: Significance test for the setup time of the 64th batch production**

The results reported above show that in all cases the two groups have different performance and such differences are not due to random effect but depend on the fact that they really belong to different populations. Aside from statistical considerations, it means that SOPHOS-MS is able to deliver a real educational advantage to people that use it for training as previously discussed when learning curves were considered for the two groups. It is interesting to notice that, just after the first lot production, the two samples do not have any overlap so the Mann-Whitney statistic U reaches its minimum value and it persists over the subsequent production levels; indeed operators' skills evolve along two separates learning curves with different learning rates.

As a further concluding remark, the system is thought also to capture relevant interaction patterns along its use to gather data and information on the operators' behaviors. Such data, handled through advanced analytics can provide meaningful insights about operators' learning curves highlighting those procedures/activities where faults are more likely to occur. In other words, the system is able to collect real time data about its usage that can be a primary input for operators' profiling and evaluation as well as for internal procedures redesign aimed at minimizing risk factors.

## CONCLUSIONS

As the Industry 4.0 takes shape, human operators experience an increased complexity of their daily tasks: they are required to be highly flexible and to demonstrate adaptive capabilities in a very dynamic working environment. Even if there has been a great deal of efforts toward smart factory concepts and engineering, such efforts are mostly related to automation systems, plant solutions, communication infrastructures, systems connectivity and interoperability and data flows management. Therefore, this research is intended to take a step forward proposing a human-centered approach along with its implementation and deployment in order to align (and enhance) operators' capabilities/competencies with the new smart factory context. The main research contribution is twofold: (i) a methodological framework aimed at shaping the augmented operator paradigm within the industry 4.0 vision; (ii) a breakthrough solution that implements the aforementioned framework. As a result, the proposed solution relies on Augmented Reality (AR) applications, that are suited to augment operators' skills and abilities to perceive and act within the working environment, but not only. Indeed, it conceptualizes, designs and implements a new approach/component to extend operators capabilities within the smart factory paradigm. It is an intelligent personal digital assistant with vocal interaction capabilities and therefore able to answer operator's questions about tasks/procedures/equipments. It is meant to provide quick and effective support to operators allowing them to acquire the information they need through a Q&A approach as they would if they were talking with a knowledgeable person.

The overall solution potentials, in terms of training effects and effects on labor performances, are investigated in a real application example concerning setup operations for cushion slides manufacturing. The evaluation has been based on the comparison of traditionally trained operators with operators trained by SOPHOS-MS over a two weeks time window. Results show that the average marginal difference between the two sets of operators grows asymptotically along the learning curves as operators gain experience demonstrating that operators trained by SOPHOS-MS outperform traditionally-trained operators all along and, as a consequence, SOPHOS-MS could be profitably used making the difference between who uses it and who does not. Such results are also supported by statistical significance tests that have been carried out in order to ascertain that the datasets related to traditionally trained operators and operators trained by SOPHOS-MS were significantly different.

**REFERENCES**

Bagheri, B., Yang, S., Kao, H., Lee, J.(2015). Cyber-physical Systems Architecture for Self-Aware Machines in Industry 4.0 Environment, *IFAC-PapersOnLine*, Volume 48, Issue 3, 2015, Pages 1622-1627, ISSN 2405-8963, http://dx.doi.org/10.1016/j.ifacol.2015.06.318.

Erol, S., Jäger, A., Hold, P., Ott, K., Sihn, W.(2016) Tangible Industry 4.0: A Scenario-Based Approach to Learning for the Future of Production, *Procedia CIRP*, Volume 54, 2016, Pages 13-18, ISSN 2212-8271, http://dx.doi.org/10.1016/j.procir.2016.03.162.

Fedorov, A., Goloschchapov, E., Ipatov, O., Potekhin, V., Shkodyrev, V., Zobnin, S. (2015) Aspects of Smart Manufacturing Via Agent-based Approach, *Procedia Engineering*, Volume 100, 2015, Pages 1572-1581, ISSN 1877-7058, http://dx.doi.org/10.1016/j.proeng.2015.01.530.

Hecklau, F., Galeitzke, M., Flachs, S., Kohl, H.(2016). Holistic Approach for Human Resource Management in Industry 4.0*, Procedia CIRP*, Volume 54, 2016, Pages 1-6, ISSN 2212-8271, http://dx.doi.org/10.1016/j.procir.2016.05.102.

Helu, M., Hedberg, T. (2015)Enabling Smart Manufacturing Research and Development using a Product Lifecycle Test Bed, *Procedia Manufacturing*, Volume 1, 2015, Pages 86-97, ISSN 2351-9789, http://dx.doi.org/10.1016/j.promfg.2015.09.066.

Helu, M.,Morris, K., Jung, K., Lyons, K., Leong, S.(2015) Identifying performance assurance challenges for smart manufacturing, *Manufacturing Letters*, Volume 6, October 2015, Pages 1-4, ISSN 2213-8463, http://dx.doi.org/10.1016/j.mfglet.2015.11.001.

Hwang, G., Lee, J., Park, J., Chang,T. (2016). Developing performance measurement system for Internet of Things and smart factory environment. International Journal Of Production Research Vol. 0 , Iss. 0,0, DOI: 10.1080/00207543.2016.1245883

Ivanov, D., Dolgui, A., Sokolov, B., Werner, F., Ivanova, M. (2016).A dynamic model and an algorithm for short-term supply chain scheduling in the smart factory industry 4.0, *International Journal Of Production Research* Vol. 54 , Iss. 2,2016

Jung, K., Choi, S., Kulvatunyou, B., Cho, H., Morris, K. C. (2016) A reference activity model for smart factory design and improvement. *Production Planning & Control* Vol. 0 , Iss. 0,0; DOI: 10.1080/09537287.2016.1237686

Kõrbe Kaare, K., Otto, T.(2015). Smart Health Care Monitoring Technologies to Improve Employee Performance in Manufacturing, *Procedia Engineering*, Volume 100, 2015, Pages 826-833, ISSN 1877-7058, http://dx.doi.org/10.1016/j.proeng.2015.01.437.

Lee, J., Bagheri, B., Kao, H. (2015). A Cyber-Physical Systems architecture for Industry 4.0-based manufacturing systems, Manufacturing Letters, Volume 3, January 2015, Pages 18-23, ISSN 2213-8463, http://dx.doi.org/10.1016/j.mfglet.2014.12.001.

Park, S. (2016) Development of Innovative Strategies for the Korean Manufacturing Industry by Use of the Connected Smart Factory (CSF), *Procedia Computer Science*, Volume 91, 2016, Pages 744-750, ISSN 1877-0509, http://dx.doi.org/10.1016/j.procs.2016.07.067.

Qin,J., Liu,Y., Grosvenor, R. (2016). A Categorical Framework of Manufacturing for Industry 4.0 and Beyond, *Procedia CIRP*, Volume 52, 2016, Pages 173-178, ISSN 2212-8271, http://dx.doi.org/10.1016/j.procir.2016.08.005.

Schumacher, A., Erol, E., Sihn, W.(2016). A Maturity Model for Assessing Industry 4.0 Readiness and Maturity of Manufacturing Enterprises, *Procedia CIRP*, Volume 52, 2016, Pages 161-166, ISSN 2212-8271, http://dx.doi.org/10.1016/j.procir.2016.07.040.

Stock, T., Seliger, G.(2016). Opportunities of Sustainable Manufacturing in Industry 4.0, *Procedia CIRP*, Volume 40, 2016, Pages 536-541, ISSN 2212-8271, http://dx.doi.org/10.1016/j.procir.2016.01.129.

Towill, D.R., Cherrington, J.E. (1994). Learning curve models for predicting the performance of advanced manufacturing technology. *International Journal of Advanced Manufacturing Technology*, 9(3), 195-203.doi.1007/BF01754598

Weyer, S., Schmitt, M., Ohmer, M., Gorecky, D.(2015). Towards Industry 4.0 - Standardization as the crucial challenge for highly modular, multi-vendor production systems*, IFAC-PapersOnLine*, Volume 48, Issue 3, 2015, Pages 579-584, ISSN 2405-8963, http://dx.doi.org/10.1016/j.ifacol.2015.06.143.